\documentclass[tighten,nofootinbib,aps,showpacs,amssymb,floatfix]{revtex4}
\newcommand{\be}{\begin{equation}}
\newcommand{\ee}{\end{equation}}
\newcommand{\beq}{\begin{eqnarray}}
\newcommand{\eeq}{\end{eqnarray}}

\usepackage{bm}
\usepackage{graphicx}%
\usepackage{epsf}
\usepackage{epsfig}

 \usepackage{amsmath}
\usepackage{amsfonts,amssymb}

\usepackage{dsfont}
\usepackage{slashed}
\usepackage{booktabs}

\begin{document}

\title{Moments of nucleon generalized  parton distributions from lattice QCD\\
\begin{figure}[h]
\begin{center}
\includegraphics[scale=0.13]{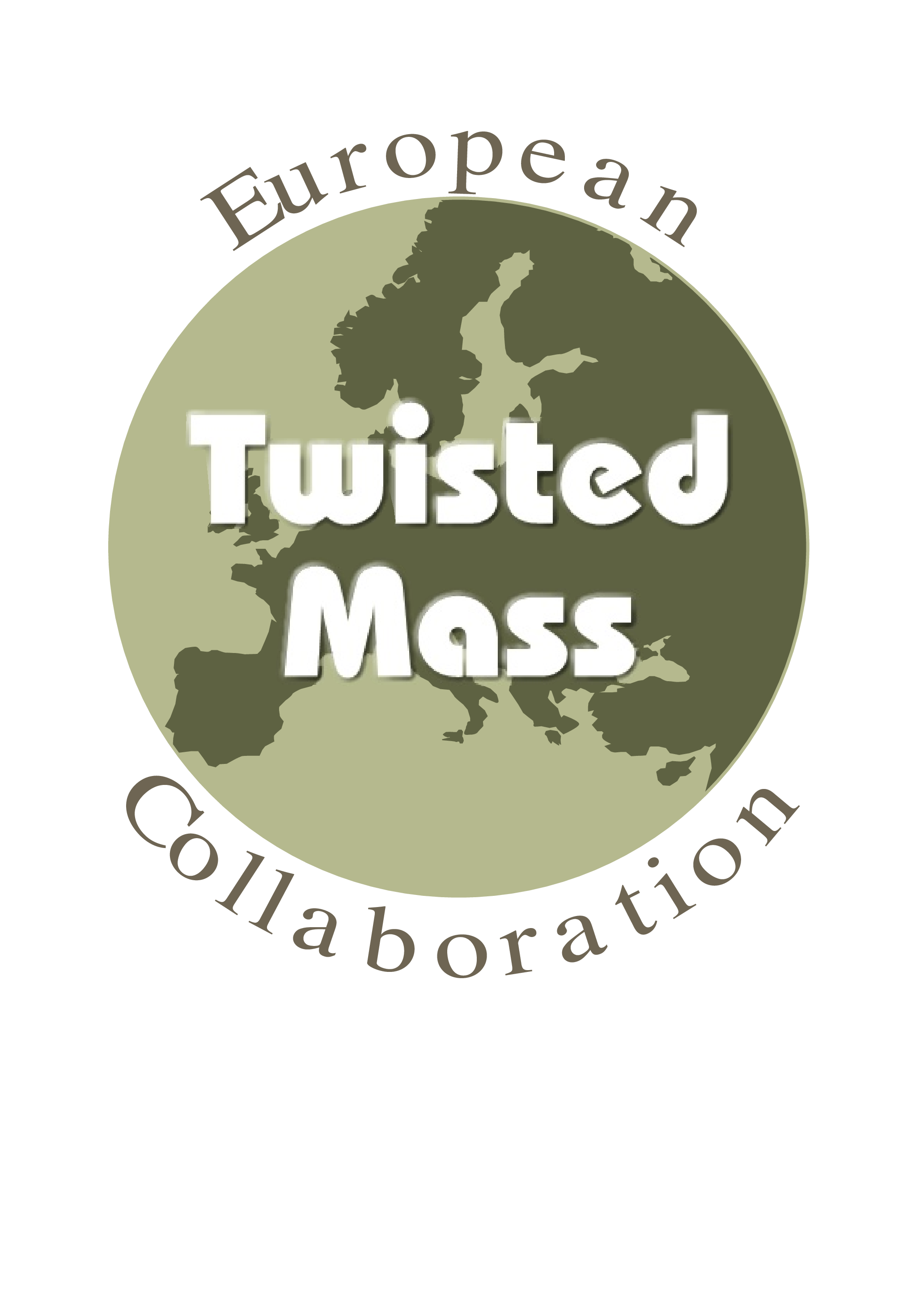}
\end{center}
\end{figure}}

\vskip -2cm
\author{\vskip -2cm C.~Alexandrou~$^{(a,b)}$, J.~Carbonell~$^{(c)}$, M. Constantinou~$^{(a)}$,  P.~A.~Harraud~$^{(c)}$,
P.~Guichon~$^{(d)}$,
K.~Jansen~$^{e}$, C. Kallidonis~$^{(a)}$, T.~Korzec~$^{(a,f)}$, M.~Papinutto~$^{(c)}$ }
\affiliation{$^{(a)}$ Department of Physics, University of Cyprus, P.O. Box 20537, 1678 Nicosia, Cyprus\\
 $^{(b)}$ Computation-based Science and Technology Research
    Center, Cyprus Institute, 20 Kavafi Str., 2121 Nicosia, Cyprus \\
$^{(c)}$ Laboratoire de Physique Subatomique et Cosmologie,
               UJF/CNRS/IN2P3, 53 avenue des Martyrs, 38026 Grenoble, France\\
$^{(d)}$ CEA-Saclay, IRFU/Service de Physique Nucl\'eaire, 91191 Gif-sur-Yvette, France\\
$^{(e)}$ NIC, DESY, Platanenallee 6, D-15738 Zeuthen, Germany\\
         \vspace{0.2cm}
 $^{(f)}$ Institut f\"ur Physik
   Humboldt Universit\"at zu Berlin, Newtonstrasse 15, 12489 Berlin, Germany}

\begin{abstract}

We present results on the lower moments of the nucleon generalized parton distributions within
lattice QCD using two dynamical flavors of degenerate twisted mass fermions. Our
simulations are performed on lattices with three different values of
the lattice spacings, namely $a=0.089$~fm, $a=0.070$~fm and
$a=0.056$~fm, allowing the investigation of cut-off effects. The volume
dependence is examined using simulations on two lattices of spatial
length $L=2.1$~fm and $L=2.8$~fm. The simulations
span pion masses in the range
of 260-470 MeV. Our results are renormalized
non-perturbatively and the values are given in the $\overline{\rm MS}$ scheme at a
scale $ \mu=2$~GeV. They are  chirally extrapolated to the physical
point in order to compare with experiment. The consequences of these results on the spin carried by the quarks in the nucleon are investigated.

\end{abstract}

\pacs{11.15.Ha, 12.38.Gc, 12.38.Aw, 12.38.-t, 14.70.Dj}

\maketitle

\setcounter{figure}{\arabic{figure}}

\newcommand{\Op}{\mathcal{O}} 
\newcommand{\PO}{\mathcal{P}} 
\newcommand{\C}{\mathcal{C}} 
\newcommand{\eins}{\mathds{1}} 
\newcommand{\J}{\mathcal{J}}
\newcommand{\Dlr}{\buildrel \leftrightarrow \over D\raise-1pt\hbox{}}

\newcommand{\twopt}[5]{\langle G_{#1}^{#2}(#3;\mathbf{#4};\Gamma_{#5})\rangle}
\newcommand{\threept}[7]{\langle G_{#1}^{#2}(#3,#4;\mathbf{#5},\mathbf{#6};\Gamm
a_{#7})\rangle}

\bibliographystyle{apsrev}                     

\section{Introduction}
Lattice QCD calculations of observables, related to the structure of
baryons, are now  carried out using simulations of the
theory with parameters that
are close enough to their physical values
that a connection of lattice results to experiment is facilitated.
This is due to the fact that  systematic
uncertainties caused by a finite volume, a finite lattice cut-off and
unphysically high pion masses are becoming better controlled. Nowadays,
a number of major collaborations are producing results on nucleon form
factors and the first moments of structure functions close to the
physical regime both in terms of pion mass and with respect to the
continuum limit~\cite{Hagler:2007xi, Syritsyn:2009mx, Brommel:2007sb,
  Alexandrou:2010cm, Yamazaki:2009zq, Alexandrou:2010hf}.
\par The Generalized
Parton Distributions (GPDs) encode important information related to
baryon structure~\cite{Mueller:1998fv,Ji:1996nm, Radyushkin:1997ki}. They occur in several physical processes such as Deeply Virtual Compton Scattering and Deeply Virtual Meson Production. Their forward limit coincides with the usual parton distributions and their first moments are related to the nucleon elastic form factors. Moreover a combination of their second moments, known as Ji's sum rule~\cite{Ji:1996ek}, allows to determine the contribution of a specific parton to the nucleon angular momentum. In the context of the ``proton spin puzzle'' this has triggered an intense experimental  activity~\cite{Chekanov:2008vy,Aaron:2007cz,Airapetian:2009rj,MunozCamacho:2006hx,Stepanyan:2001sm}.

The GPDs can be accessed in high energy processes  where QCD factorization applies, in which case the amplitude is the convolution of a hard perturbative kernel with the GPDs, as illustrated in Fig.~\ref{handbag}. Generically the GPDs are defined as matrix elements of bi-local operators separated by a light-like interval. Due to the Wick rotation such matrix elements cannot be computed directly on the Euclidean lattice. Instead one considers their 
Mellin moments, which in principle, carry the same information.
\par If   $|p^\prime
\rangle $ and $|p\rangle$ are one-particle states, the twist-2 GPDs, which are studied in this paper, are defined by the matrix element~\cite{Diehl:2003ny}:

\beq
  F_{\Gamma}(x,\xi,q^2) =   \frac{1}{2}\int \! \frac{d\lambda}{2\pi}
     e^{ix\lambda} \langle p^\prime |\bar {\psi}(-\lambda n/2) {\Gamma}
\PO      e^{ig\!\int \limits_{-\lambda /2}^{\lambda /2}\! d\alpha\, n \cdot A(n\,\alpha)}
     \psi(\lambda n/2) |p\rangle \,,\label{bilocal}
\eeq
where $q=p^\prime-p$, $\xi=-n\cdot q/2$, $x$ is the
momentum fraction, and $n$ is a light-like  vector collinear to $P=(p+p')/2$ and such that
${P}\cdot n=1$. The gauge link $\PO \exp(\dots) $  is necessary for gauge invariance. In model calculations it is often set to one, which amounts to
working with QCD in the light-like gauge
$A\cdot n=0$,  but on  the lattice such a gauge fixing is not necessary.
\par In this work we  shall consider only the GPDs corresponding to a Dirac structure $\Gamma$  which conserves the quark chirality, that is $\Gamma=\slashed{n}$ and  $\Gamma=\slashed{n}\gamma_5$. 
The associated matrix elements may be parametrized in the following way:
\begin{figure}[h]
 \includegraphics[scale=0.8]{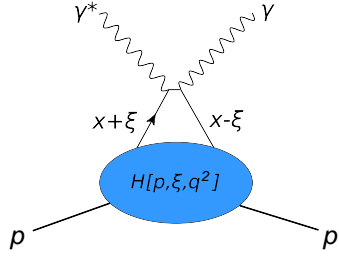}
\caption{``Handbag'' diagram.}
\label{handbag}
\end{figure}
%
\beq
 F_{\slashed{n}}(x,\xi,q^2) &=&\ \frac{1}{2}\bar{u}_N(p^\prime)\left[\slashed{n}{ H(x,\xi,q^2)}+i\frac{n_\mu q_\nu \sigma^{\mu\nu}}{2m_N}{ E(x,\xi,q^2)}\right] u_N(p) \\
F_{\slashed{n}\gamma_5}(x,\xi,q^2) & =& \frac{1}{2}\bar{u}_N(p^\prime)\left[\slashed{n}\gamma_5 {{\tilde{H}(x,\xi,q^2)}}+\frac{n\cdot q \gamma_5}{2m_N} {\tilde{E}(x,\xi,q^2)}\right] u_N(p).\label{twist2GPD}
\eeq
where  $u_N$ is a nucleon spinor and  $H,E,\tilde{H},\tilde{E}$ are the twist-2 chirality even  GPDs.
In the forward limit for which  $\xi=0$ and $q^2=0$ they reduce to  the ordinary parton distributions, namely the longitudinal momentum  $q(x)$ and the
helicity $\Delta q(x)$ distributions are given by: 
\begin{equation}
   q(x) = H(x,0,0), \qquad {\rm and} \qquad \Delta q(x) = \tilde H(x,0,0) \, .
\end{equation}
The first few Mellin moments of these parton distributions are of particular interest
\begin{eqnarray}
   \langle x^{n-1}\rangle_q          &=& \int_{-1}^1 x^{n-1} q(x)\ dx \\
   \langle x^{n-1}\rangle_{\Delta q} &=& \int_{-1}^1 x^{n-1} \Delta q(x)\ dx \, .
\end{eqnarray}
Since as already mentioned, matrix elements of the light-cone operator as defined in~Eq.(\ref{bilocal}) cannot be extracted from 
correlators in euclidean lattice QCD, the usual method is to proceed with an operator product
expansion of this operator that leads to a tower of local operators given by:
\beq
\vspace{-1cm}
   \Op_V^{\mu_1\ldots\mu_{n}}    &=&
  \bar \psi  \gamma^{\{\mu_1}i\Dlr^{\mu_2}\ldots i\Dlr^{\mu_{n}\}} \psi  \\
\Op_{A}^{\mu_1\ldots\mu_{n}}      &=&
\bar \psi  \gamma^{\{\mu_1}i\Dlr^{\mu_2}\ldots i\Dlr^{\mu_{n}\}}\gamma_5 \psi   \, .
\label{Oper_def}\eeq
%
%
The curly brackets represent a symmetrization over indices and
subtraction of traces. The computation of the
 matrix elements of these operators on the Euclidean lattice can be
done with standard techniques. The case $n=1$ amounts to calculating the elastic form factors  of the vector and axial-vector  currents and the results are reported in
Ref.~\cite{Alexandrou:2010cm,Alexandrou:2011db}. In this work we
concentrate on the $n=2$ moments, i.e. matrix elements of operators with a single derivative.
The matrix elements of the these operators are  parametrized in terms of
  the generalized form
factors (GFFs) $A_{20}(q^2),\ B_{20}(q^2),\ C_{20}(q^2)$ and $\tilde
A_{20}(q^2),\ \tilde B_{20}(q^2)$, according to
\beq
  \langle N(p^\prime,s^\prime)| \Op_\slashed{n}^{\mu\nu}| N(p,s)\rangle &=&
   \bar u_N(p^\prime,s^\prime)\Bigl[ A_{20}(q^2)\, \gamma^{\{ \mu}P^{\nu\}}
                      +B_{20}(q^2)\, \frac{i\sigma^{\{
                          \mu\alpha}q_{\alpha}P^{\nu\}}}{2m}
                      +C_{20}(q^2)\, \frac{1}{m}q^{\{\mu}q^{\nu\}} \Bigr]
   u_N(p,s)\, , \nonumber \\
\label{eq_decomposition}\\
  \langle N(p^\prime,s^\prime)| \Op_{\slashed{n}\gamma_5}^{\mu\nu}| N(p,s)\rangle &=&
 \bar u_N(p^\prime,s^\prime)\Bigl[ \tilde A_{20}(q^2)\, \gamma^{\{
     \mu}P^{\nu\}}\gamma^5
                      + \tilde B_{20}(q^2)\,
                      \frac{q^{\{\mu}P^{\nu\}}}{2m}\gamma^5\Bigr]
 u_N(p,s)\, .\nonumber
\eeq

Note that the GFFs depend only on the squared momentum transfer
 $q^2=(p^\prime-p)^2$ which   implies that the moments of the GPDs are  polynomial in  $\xi$.
In the forward limit we have
$A_{20}(0)=\langle x \rangle_q $ and $\tilde A_{20}(0)=\langle
x \rangle_{\Delta q} $, which are respectively the first moment of the unpolarized and polarized quark distributions.
Knowing the GFFs  one can  evaluate the
quark contribution to the nucleon spin using  Ji's sum rule: $J^q = \frac{1}{2}[ A_{20}^q(0) +
  B_{20}^q(0)]$. Moreover, using the measured or calculated value of the quark helicity $\Delta \Sigma^q=g_A^q $ the decomposition   $J^q=\frac{1}{2}\Delta \Sigma^q+ L^q$  allows to study the role  of the quark orbital angular momentum $L^q$.

\section{Lattice evaluation}

Twisted mass fermions~\cite{Frezzotti:2000nk}  provide an attractive  formulation of lattice QCD that
allows for automatic ${\cal O}(a)$ improvement, infrared regularization
of small
eigenvalues and fast dynamical
simulations~\cite{Frezzotti:2003ni}. For the
calculation of the moments of
GPDs, which is the main focus of this work, the automatic
 ${\cal O}(a)$ improvement is particularly relevant
since it is achieved  by tuning only one parameter in the action,
requiring no further improvements on the operator level.

The action for two degenerate flavors of quarks
 in twisted mass QCD is given by
   \begin{equation}
      S=S_g + \sum_x \bar\chi(x) \left[D_W
 {+} m_{\rm crit}
 {+} i\gamma_5\tau^3\mu \right]\chi(x)\,,
   \end{equation}
where $D_W$ is the Wilson Dirac operator and we use the tree-level Symanzik improved
gauge action $S_g$~\cite{Weisz:1982zw}. The quark fields $\chi$
are in the so-called ``twisted basis''
obtained from the ``physical basis''
at
maximal twist by a simple transformation:
\be
\psi {=} \frac{1}{\sqrt{2}}[{\bf 1} + i\tau^3\gamma_5]\chi \quad {\rm and}
\quad \bar\psi {=} \bar\chi \frac{1}{\sqrt{2}}[{\bf 1} + i\tau^3\gamma_5]\,.
\ee
We note that, in the continuum, this action is equivalent to the standard
QCD action. As we pointed out, a
crucial advantage is the fact that by tuning a single parameter,
namely the bare untwisted quark mass to its critical value $m_{\rm
  crit}$,
a wide class of physical observables are automatically ${\cal O}(a)$
improved~\cite{Frezzotti:2000nk,Frezzotti:2003ni,Shindler:2007vp}. A disadvantage is  the explicit flavor symmetry breaking. In
a recent paper we have checked that this breaking is small for the baryon
observables under consideration in this work and for the lattice spacings
that we use~\cite{Alexandrou:2009xk,Alexandrou:2009qu,Drach:2009dh,Alexandrou:2008tn,Alexandrou:2007qq}. To
extract the GFFs without needing to evaluate
the disconnected contributions we evaluate the  nucleon matrix
elements  corresponding to the operators defined by
\begin{equation}\label{eq_operators}
\Op_{V^a}^{\mu_1 \ldots \mu_n} = \bar \psi\,\gamma^{\{\mu_1} iD^{\mu_2}\ldots iD^{\mu_{n}\}} \frac{\tau^a}{2} \, \psi,
\qquad  \Op_{A^a}^{\mu_1\ldots\mu_n} = \bar \psi\,\gamma^5\gamma^{\{\mu_1}  iD^{\mu_2}\ldots iD^{\mu_{n}\}}\frac{\tau^a}{2} \, \psi \, ,
\end{equation}
where from now on we use the notation $\Op_{V^a}^{\mu \ldots \mu_n}$ and $\Op_{A^a}^{\mu \ldots \mu_n}$ to denote the vector and axial-vector operators
with  flavor index $a$.
These matrix elements
receive contributions only from the connected diagram for $a=1,\,2$ and up to ${\cal O}(a^2)$ for $a=3$~\cite{Dimopoulos:2009qv}. In particular, we consider
the isovector combination with $a=3$ for which the form of the operators
remain the same in the physical and twisted basis.
In order to find the spin carried by each quark in the nucleon we
also analyse the isoscalar one-derivative vector and axial-vector operators.
The latter receive contributions from disconnected fermion loops, which
we neglect in this analysis.
Simulations including a dynamical strange quark are also available within the
twisted mass formulation. Comparison of the nucleon mass obtained with two
dynamical flavors and the nucleon mass including a dynamical strange quark
has shown negligible dependence on the dynamical strange quark~\cite{Drach:2010}.
We therefore expect the results on the nucleon moments to show little
sensitivity on a dynamical strange quark. This is also confirmed by
comparing our results to those where a dynamical strange quark is included.

In this work we consider simulations at three values of the coupling constant
with lattice spacings 0.056~fm, 0.07~fm and to 0.089~fm determined
from the nucleon mass. This enables us to
obtain results in the continuum limit.  We also examine finite size effects
 by comparing results on two lattices of spatial length $L=2.1$~fm and $L=2.8$~fm~\cite{Alexandrou:2010,Alexandrou:2009ng,Alexandrou:2008rp}.

\subsection{Correlation functions}

The GFFs are extracted from dimensionless ratios of
correlation functions. The two-point and three-point functions are
defined by
\beq
G(\vec q, t_f-t_i)\hspace{-0.15cm}&=&\hspace{-0.25cm}\sum_{\vec x_f} \, e^{-i(\vec x_f-\vec x_i) \cdot \vec q}\,
     {\Gamma^0_{\beta\alpha}}\, \langle {J_{\alpha}(t_f,\vec x_f)}{\overline{J}_{\beta}(t_i,\vec{x}_i)} \rangle \label{twop}\\
G^{\mu_1 \cdots\mu_n}(\Gamma^\nu,\vec q, t-t_i) \hspace{-0.15cm}&=&\hspace{-0.25cm}\sum_{\vec x, \vec x_f} \, e^{i(\vec x -\vec x_i)\cdot \vec q}\,  \Gamma^\nu_{\beta\alpha}\, \langle
{J_{\alpha}(t_f,\vec x_f)} \Op^{\mu_1 \cdots \mu_n}(t,\vec x) {\overline{J}_{\beta}(t_i,\vec{x}_i)}\rangle\,,
\eeq
where we consider kinematics for which the final momentum  $p_f=0$. We 
 drop $t_f-t_i$ from the argument of the three-point function since it will be
kept fixed in our approach. The 
projection matrices ${\Gamma^0}$ and ${\Gamma^k}$ are given by:
\be
{\Gamma^0} = \frac{1}{4}(\eins + \gamma_0)\,,\quad {\Gamma^k} =
i{\Gamma^0} \gamma_5 \gamma_k\,\quad k=1,2,3.\label{threep}
\ee
 The proton interpolating field written in the twisted basis at
 maximal twist is given by
\be
\tilde{J}(x) {=} {\frac{1}{\sqrt{2}}[\eins + i\gamma_5]}\epsilon^{abc} \left[ {\tilde{u}}^{a \top}(x) \C\gamma_5 \tilde{d}^b(x)\right] {\tilde{u}}^c(x).
\ee
where $\C$ is the charge conjugation matrix. We use Gaussian smeared
quark fields~\cite{Alexandrou:1992ti,Gusken:1989} to increase
the overlap with the proton state and decrease overlap with excited
states. The smeared interpolating fields are given by:
\beq
q_{\rm smear}^a(t,\vec x) &=& \sum_{\vec y} F^{ab}(\vec x,\vec y;U(t))\ q^b(t,\vec y)\,,\\
F &=& (\eins + {\alpha} H)^{n} \,, \nonumber\\
H(\vec x,\vec y; U(t)) &=& \sum_{i=1}^3[U_i(x) \delta_{x,y-\hat\imath} + U_i^\dagger(x-\hat\imath) \delta_{x,y+\hat\imath}]\,. \nonumber
\eeq
We also apply APE-smearing to the gauge fields $U_\mu$ entering
the hopping matrix $H$. The parameters for the Gaussian smearing $\alpha$ and $n$ are optimized using the nucleon
mass as described in Ref.~\cite{Alexandrou:2010hf}.

\begin{figure}
 \includegraphics[scale=0.75]{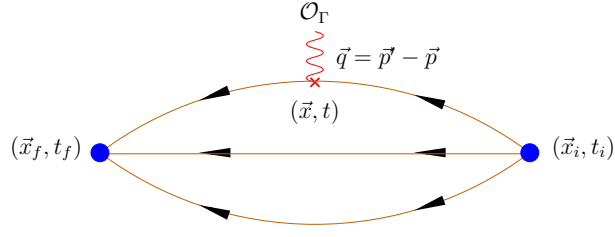}
\caption{Connected nucleon three-point function.}
\label{fig:connected_diagram}
\end{figure}

For correlators containing the
isovector operators the disconnected diagrams are zero up to
lattice artifacts, and can be safely neglected as we approach the
continuum limit. The detailed investigation of volume and cut-off effects
will be performed on isovector quantities, for which no contributions
are neglected. They can be calculated by evaluating the
connected diagram, shown 
schematically in
Fig.~\ref{fig:connected_diagram}. A standard approach
to calculate the connected three-point function is using sequential
inversions through the sink~\cite{Dolgov:2002zm}.
The creation  operator is taken at
at a fixed position  $\vec{x}_i {=} {\vec 0}$ (source). The
annihilation operator at a later time $t_f$ (sink) carries momentum
${\bf p}^\prime {=} 0$. The current couples to a quark at an
intermediate time $t$ and carries momentum ${\bf q}$. Translation
invariance enforces ${\bf q}=-{\bf p}$ for our kinematics.
At a fixed source-sink time separation we
obtain results for all possible momentum transfers and
insertion times as well as for any operator $\Op^{\{\mu_1\cdots\mu_n\}}$, with
one set of sequential inversions per choice of the sink.
We perform separate inversions for each one of the four projection matrices
$\Gamma^\mu$ given in Eq.~(\ref{threep}).

\begin{figure}[h]
   \includegraphics[width=0.49\linewidth]{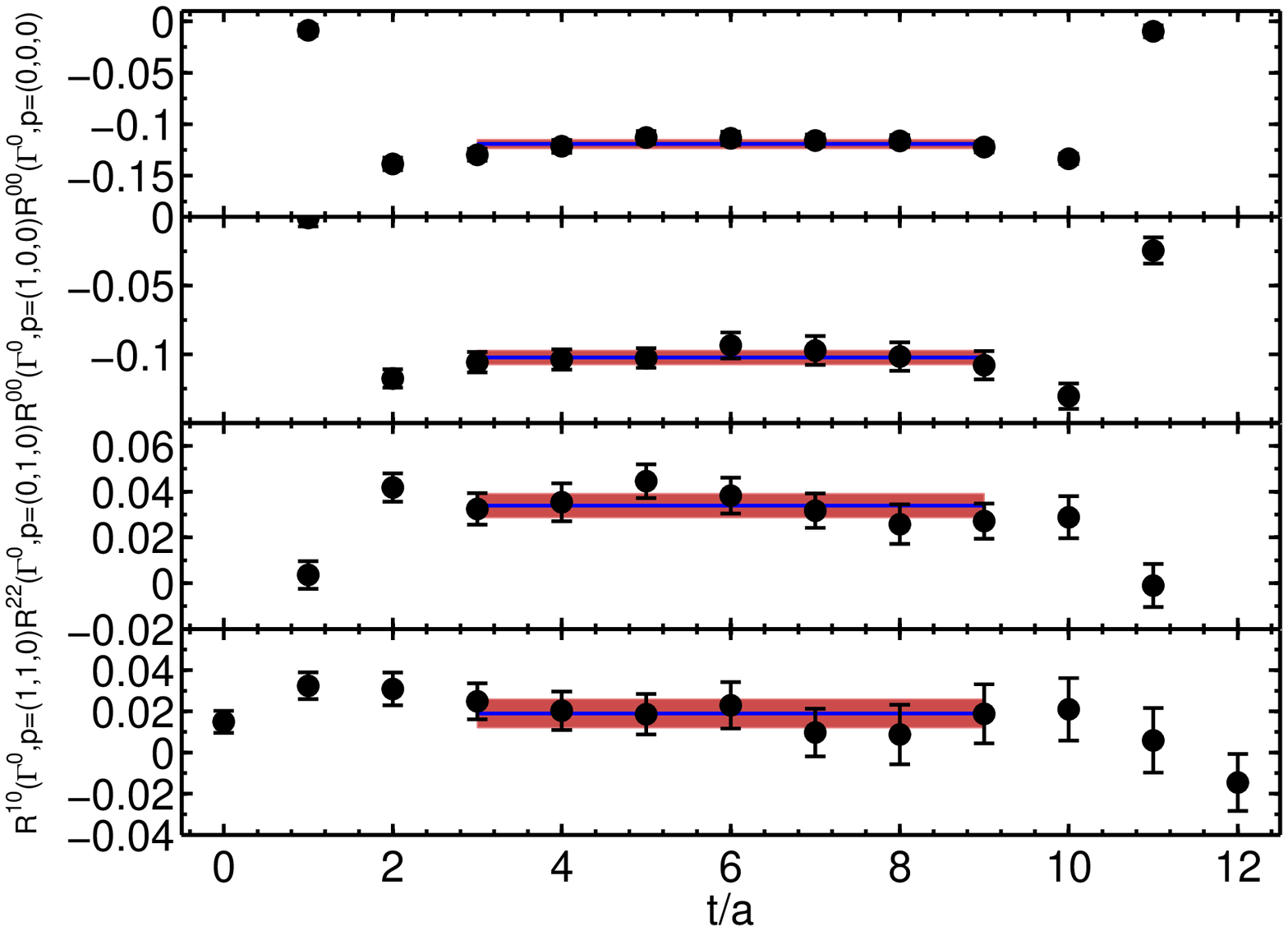}\includegraphics[width=0.49\linewidth]{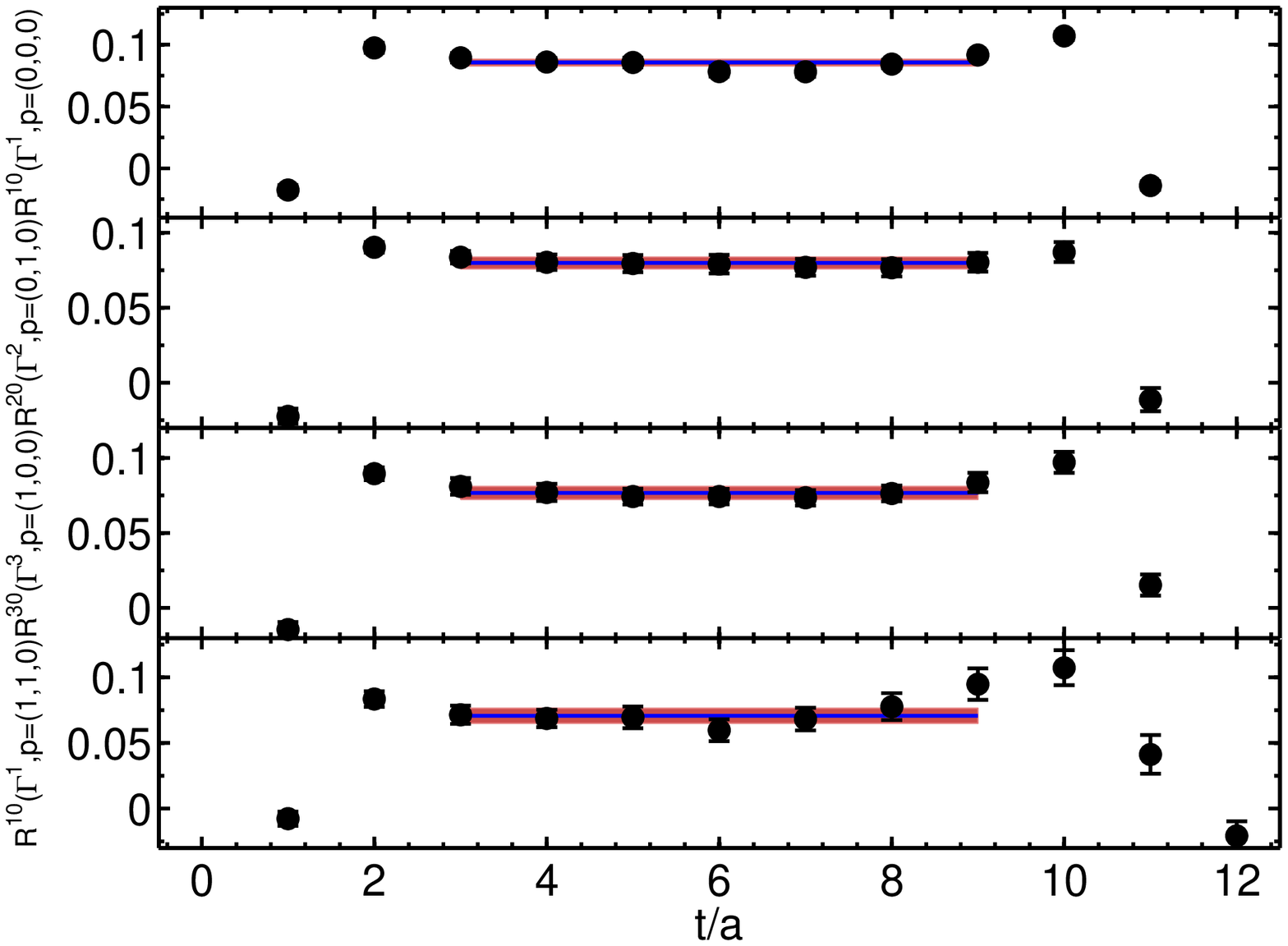}
   \caption{\label{fig:plateaus} Ratios for the one derivative vector (left)
            and axial vector (right) operator for a few exemplary choices of the momentum. The solid lines with the bands
            indicate the fitted plateau values with their jackknife errors.
From top to bottom the  momentum takes values $\vec{p}=(0,0,0); (0,1,0), (1,0,0)$ and $(1,1,0)$.}
\end{figure}

Using the two- and three-point functions of
Eqs.~(\ref{twop})-(\ref{threep}) and considering only
one derivative operators we form the ratio
\be
R^{\mu\nu}(\Gamma^\lambda,\vec q,t)= \frac{G^{\mu\nu}(\Gamma^\lambda\vec q,t) }{G(\vec 0,
  t_f)}\ \sqrt{\frac{G(\vec p, t_f-t)G(\vec 0,  t)G(\vec0,
    t_f)}{G(\vec 0  , t_f-t)G(\vec p,t)G(\vec p,t_f)}}\,,
\label{ratio}
\ee
which is optimized because it does not contain potentially noisy
two-point functions at large separations and because correlations
between its different factors reduce the statistical noise.
For sufficiently large separations $t_f-t$ and $t-t_i$ this ratio
becomes time-independent (plateau region):
\be
\lim_{t_f-t\rightarrow \infty}\lim_{t-t_i\rightarrow \infty}R^{\mu\nu}(\Gamma^\lambda,\vec q,t)=\Pi^{\mu\nu}(\Gamma^\lambda,\vec q) \,.
\label{plateau}
\ee
From the plateau values of the renormalized
asymptotic ratio $\Pi(\Gamma^\lambda, \vec q)_R=Z\Pi(\Gamma^\lambda, \vec{q})$
  the generalized form
factors can be extracted. The equations relating  $\Pi(\Gamma^\lambda,\vec q)$
to the GFFs are given in Appendix A.
All values of $\vec q$ resulting in the same
$q^2$, the four choices of $\Gamma$ and the ten orientations $\mu,\nu$
of the operator lead to an over-constrained system of equations which
is solved in the least-squares sense via a singular value
decomposition of the coefficient matrix.
All quantities will be given in Euclidean space
with $Q^2=-q^2$ the Euclidean momentum transfer squared.
 The coefficients follow from
the matrix-element decomposition given in Eq.~(\ref{eq_decomposition})
and may depend on the energy and mass of the nucleon as well as on the
initial spatial momentum $\vec p=-\vec q$ (see Appendix A). It turns out that both
the operators with $\mu=\nu$ and $\mu\neq \nu$ are necessary to obtain
all three one-derivative vector form factors. Since those two classes
of operators on a lattice renormalize differently from each
other~\cite{Gockeler:1996mu}, renormalization has to be carried out
already on the level of the ratios. Although  the one-derivative
axial form factors can be extracted using only correlators with
$\mu\neq \nu$, we use all combinations of $\mu,\nu$ in order to
decrease the statistical error. In Fig.~\ref{fig:plateaus} we show
representative  plateaus for different momentum and indices $\mu$ and $\nu$.

Since we use sequential inversions through the sink we need to fix the
sink-source separation. The statistical errors on the
three-point function are kept as small as
possible by using the smallest  value for the sink-source time separation
that still ensures that the
excited state contributions are sufficiently suppressed.
  We have tested different values of the
sink-source time separation~\cite{Alexandrou:2011db} and for the final
results we use the following values which correspond to $t_f-t_i\sim1$ fm
\be
\beta=3.9:\,(t_f-t_i)/a{=}12\,,\quad\beta=4.05:\,(t_f-t_i)/a{=}16\,,\quad\beta=4.20:\,(t_f-t_i)/a{=}18. \nonumber
\ee

\subsection{Simulation details}

The input parameters of the calculation, namely $\beta$, $L/a$ and $a\mu$
are summarized in Table~\ref{Table:params}. The  lattice spacing $a$
is determined from the nucleon mass, and the reader is referred to
Ref.~\cite{Alexandrou:2010hf,Alexandrou:2011db} for more details. Here
we present only the final values, which are
\be
a_{\beta=3.9}=0.089(1)(5)\,, \quad
a_{\beta=4.05}=0.070(1)(4)\,,\quad
a_{\beta=4.2}=0.056(2)(3)\,,\nonumber
\ee
where the first error is statistical and the second systematic.
The pion mass values, spanning a  range from 260~MeV to 470~MeV, are
taken from Ref.~\cite{Urbach:2007}. At $m_{\pi}\approx 300$ MeV and
$\beta{=}3.9$ we have simulations for lattices of spatial size
$L{=}2.1$~fm and $L{=}2.8$~fm allowing to investigate finite size
effects. Finite lattice spacing effects are studied using three sets
of results at $\beta{=}3.9$, $\beta{=}4.05$ and $\beta{=}4.2$ for the
lowest and largest pion mass available in this work. These sets of
gauge ensembles allow us to estimate all the systematic errors in
order to produce reliable predictions for the nucleon one-derivative
GFFs.

{\small{
\begin{center}
\begin{table}[h]
\begin{tabular}{c|llllll}
\hline\hline
\multicolumn{6}{c}{$\beta=3.9$, $a=0.089(1)(5)$~fm,   ${r_0/a}=5.22(2)$}\\\hline
$24^3\times 48$, $L=2.1$~fm &$a\mu$         &   &    0.0040      &   0.0064     &  0.0085     &   0.010 \\
                               & No. of confs &  &943 &553 & 365 &477 \\
                               &$m_\pi$~(GeV) &  & 0.3032(16) & 0.3770(9) & 0.4319(12) & 0.4675(12)\\
                               &$m_\pi L$     &    & 3.27       & 4.06      & 4.66       & 5.04     \\
$32^3\times 64$, $L=2.8$~fm  &$a\mu$ & 0.003 & 0.004 & & & \\
                               & No. of confs & 667  &351 & & & \\
                               & $m_\pi$~(GeV)& 0.2600(9)   & 0.2978(6) & & &  \\
                               & $m_\pi L$    & 3.74        & 4.28      &&& \\\hline \hline
\multicolumn{6}{c}{ $\beta=4.05$, $a=0.070(1)(4)$~fm, ${r_0/a}=6.61(3)$ }\\
\hline
$32^3\times 64$, $L=2.13$~fm &$a\mu$         & 0.0030     & 0.0060     & 0.0080     & \\
                               & No. of confs   &447 &326 &  419 &\\
                               &$m_\pi$~(GeV) & 0.2925(18) & 0.4035(18) & 0.4653(15) &  \\
                               &$m_\pi L$     & 3.32       &   4.58     & 5.28       &       \\ \hline\hline
\multicolumn{6}{c}{ $\beta=4.2$, $a=0.056(1)(4)$~fm  ${r_0/a}=8.31$}\\\hline
$32^3\times 64$, $L=2.39$~fm &$a\mu$          & 0.0065     &     & \\
                               & No. of confs       & 357 &  & & \\
                               &$m_\pi$~(GeV) & 0.4698(18) &  &  \\
                               &$m_\pi L$     & 4.24       &      & \\
$48^3\times 96$, $L=2.39$~fm &$a\mu$         & 0.002      &      &     & \\
                               & No. of confs      & 245          &  &  & & \\
                               &$m_\pi$~(GeV) & 0.2622(11) &  &  &  \\
                               &$m_\pi L$     & 3.55       &  &      &       \\ \hline

\end{tabular}
\caption{Input parameters ($\beta,L,a\mu$) of our lattice calculation and corresponding lattice spacing ($a$) and pion mass ($m_{\pi}$).}
\label{Table:params}
\vspace*{-.0cm}
\end{table}
\end{center}
}}

\subsection{Renormalization}
We determine the renormalization constants for the one-derivative
operators non-perturbatively, in the RI'-MOM scheme~\cite{Alexandrou:2010me}.
We employ a momentum
source~\cite{Gockeler:1998ye} and perform a perturbative subtraction
of ${\cal O}(a^2)$
terms~\cite{Constantinou:2009tr,Alexandrou:2010me}. This subtracts the
leading cut-off effects yielding  only a very weak dependence of the
renormalization factors on $(ap)^2$ for which the $(ap)^2\rightarrow
0$ limit can be reliably taken. It was also shown with high accuracy
that the quark mass dependence is negligible for the aforementioned
operators. We find the values
\beq
Z_V^{\mu=\nu}&=&0.970(26)\,,\,\, 1.013(14)\,,\,\,1.097(6)\nonumber \\
Z_V^{\mu\ne\nu}&=&1.061(29)\,,\,\, 1.131(18)\,,\,\,1.122(10) \\
Z_A^{\mu\ne\nu}&=&1.076(1)\,,\,\,\, 1.136(0)\,,\,\,\,1.165(10)\nonumber
\label{renormalization}
\eeq
at $\beta{=}$3.9, 4.05 and 4.2 respectively. These are the values
that we use in this work to renormalize the lattice matrix element.

\section{Lattice results}
In this section we present our results on the nucleon generalized form
factors $A_{20}(Q^2), B_{20}(Q^2), C_{20}(Q^2)$ and $\tilde{A}_{20}(Q^2), \tilde{B}_{20}(Q^2)$. We examine their dependence on the
lattice volume and spacing, as well as, on the pion mass.
 We also compare with recent results from
other collaborations.
In particular, we discuss lattice artifacts
for the results on the isovector combination for the renormalized
nucleon matrix element of the one-derivative operators
\be
\bar u \gamma_{\{\mu} \stackrel{\leftrightarrow}{ D}_{\nu\}} u - \bar d \gamma_{\{\mu} \stackrel{\leftrightarrow}{ D}_{\nu\}} d\,,\qquad
\bar u \gamma_5\gamma_{\{\mu} \stackrel{\leftrightarrow}{
  D}_{\nu\}} u - \bar d \gamma_5\gamma_{\{\mu}
\stackrel{\leftrightarrow}{ D}_{\nu\}} d \nonumber
\ee
 in the $\overline{\rm MS}$
scheme at a scale $ \mu=2$~GeV.

In order to obtain some estimates on the spin content of the nucleon we also
analyse the isoscalar parts of the spin-independent and helicity quark
distributions, which however neglect the disconnected contributions.

The GFFs
 $A_{20}(Q^2=0)$ and $\tilde A_{20}(Q^2=0)$ are computed
directly from the matrix elements, whereas
$B_{20}(Q^2=0)$, $C_{20}(Q^2=0)$ and $\tilde B_{20}(Q^2=0)$ are
obtained by linearly extrapolating  the $Q^2\neq 0$ data.
 $C_{20}$ is
consistent to zero within error bars for all momentum transfers.

\subsection{Finite volume effects}
\begin{figure}[h]
{\includegraphics[scale=0.5]{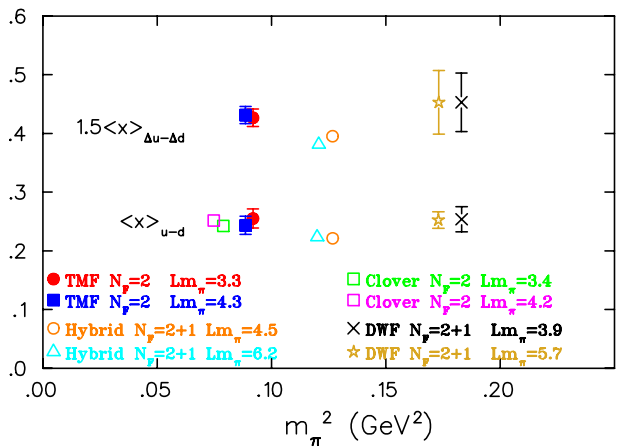}}
\caption{$\langle x\rangle_{u-d}$ and $\frac{3}{2}\langle x\rangle_{\Delta
    u-\Delta d}$ using  twisted mass fermions (this work), $N_F=2$ clover fermions~\cite{Pleiter:2011gw} (preliminary), hybrid~\cite{Bratt:2010jn} and DWF~\cite{Aoki:2010xg}.}
\label{fig:A20_A20tilde_vol}
\end{figure}

In order to access volume effects we compare in
Fig.~\ref{fig:A20_A20tilde_vol}  results on
the moments $\langle x \rangle _{u-d}$
and $\langle  x \rangle _{\Delta u-\Delta d}$ computed  on different
lattice sizes as a function of $m_\pi^2$.
As already mentioned, both these quantities are directly obtained at $Q^2=0$ and
require no assumption on their $Q^2$-dependence.
 Alongside our results we also show
results using $N_F=2$ clover fermions~\cite{Pleiter:2011gw} (preliminary), $N_F=2+1$ domain wall fermions (DWF)~\cite{Aoki:2010xg} and  domain wall valence quarks on an $N_F=2+1$ staggered sea (hybrid)~\cite{Bratt:2010jn}.

\begin{figure}[h]
{\includegraphics[scale=0.5]{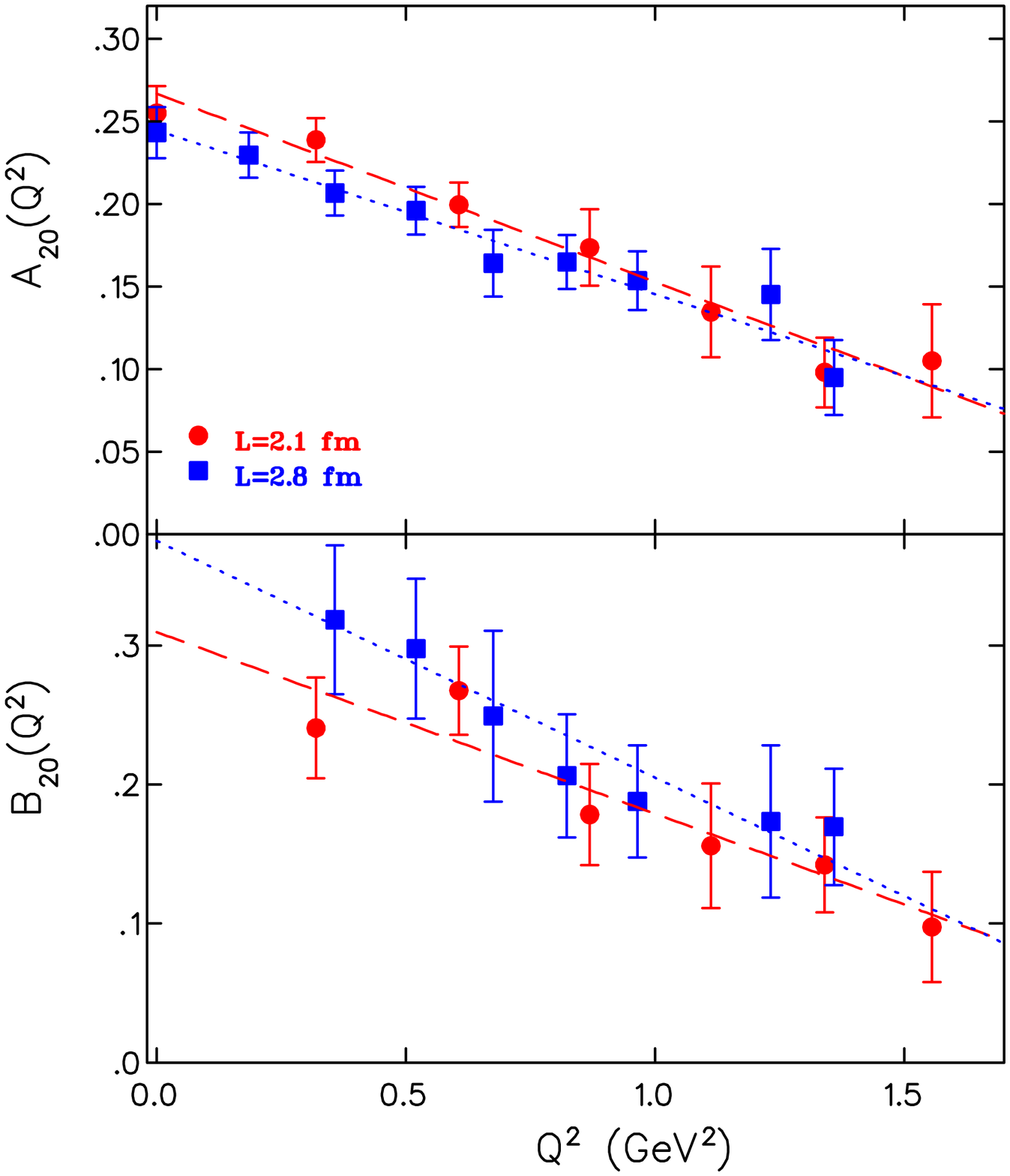}}
{\includegraphics[scale=0.5]{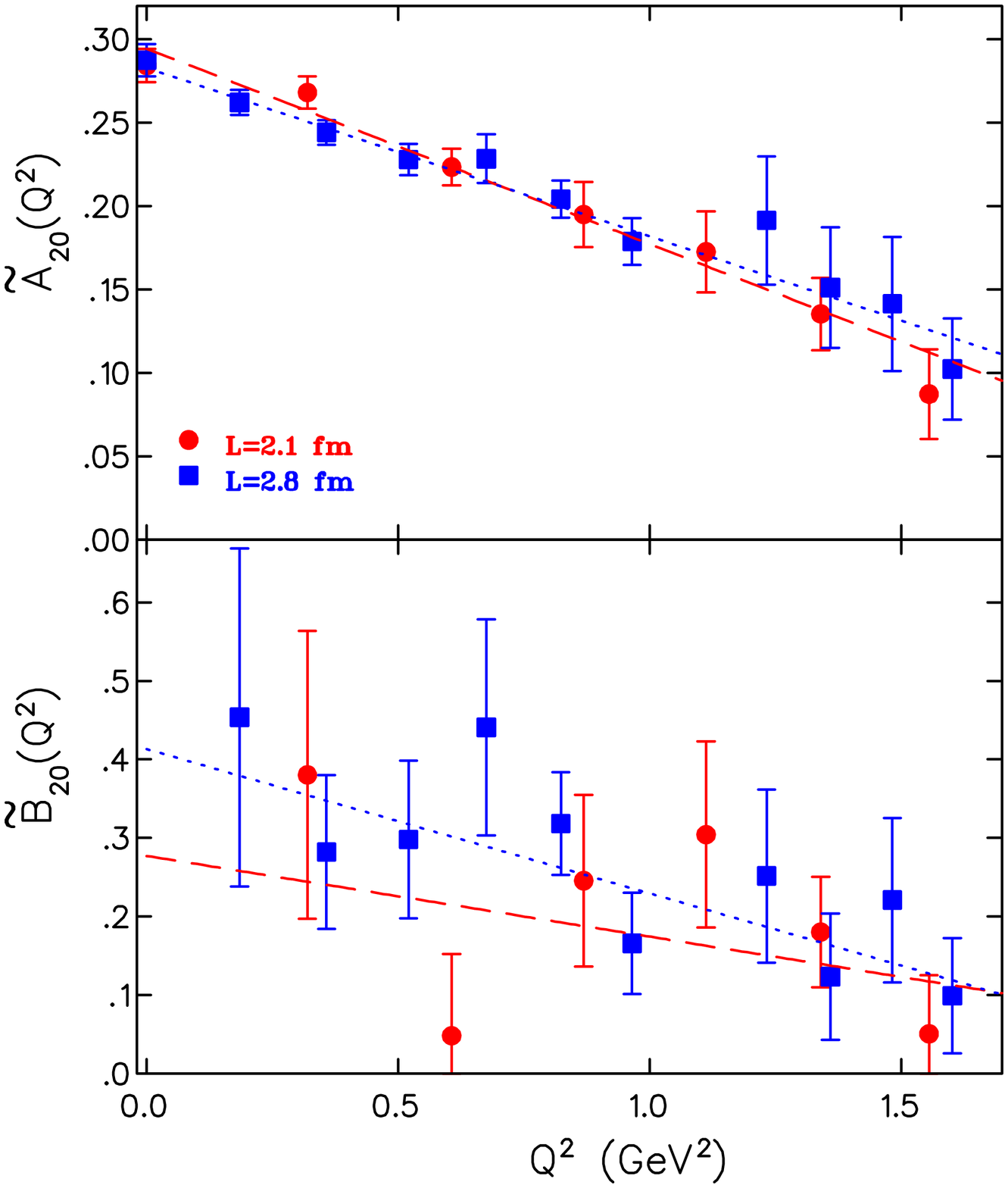}}
\caption{Left panel: The isovector GFFs $A_{20}$ and $B_{20}$ for pion
mass $\sim 300$~MeV for a lattice of spatial length $L=2.1$~fm and $L=2.8$~fm.
Right panel: The axial-vector GFFs $\tilde{A}_{20}$ and $\tilde{B}_{20}$ for
$L=2.1$~fm and $L=2.8$~fm. }
\label{fig:vol}
\end{figure}

The results shown in Fig.~\ref{fig:vol} using twisted mass fermions correspond to a pion mass
of about $ 300$~MeV and are computed on lattices of spatial $L$ with
 $Lm_\pi=3.3$ and  $Lm_\pi=4.3$. As can be seen,
results on these two lattices for both $\langle x\rangle_{u-d}$
and $\langle x\rangle_{\Delta u-\Delta d}$ are consistent.
The LHPC using
 a hybrid approach and  $m_\pi\sim
 350$~MeV has very accurate results at two lattices
  with $Lm_\pi=4.5$ and  $Lm_\pi=6.2$. No volume effects are seen
for both vector and axial-vector first moments.
The QCDSF collaboration has preliminary results for $\langle x \rangle _{u-d}$ using clover fermions  for $m_\pi \sim 270$~MeV
with $Lm_\pi=3.4$ and $Lm_\pi=4.2$, which are consistent.
Finally the   RBC-UKQCD results with domain wall fermions  with
$Lm_\pi=3.9$ and $Lm_\pi=5.7$ show no  volume
effects for both $\langle x \rangle _{u-d}$ and $\langle x\rangle_{\Delta
  u- \Delta d}$
~\cite{Aoki:2010xg}.
The conclusion that we draw from this comparison 
is that finite volume effects on $\langle x\rangle_{u-d}$ and $\langle x\rangle_{\Delta u-\Delta d}$ are insignificant at our current statistical precision for lattices that satisfy $L m_\pi > 3.3$.

In Figs.~\ref{fig:vol} we compare  results on the GFFs $A_{20}(Q^2)$, $B_{20}(Q^2)$, $\tilde{A}_{20}(Q^2)$ and $\tilde{B}_{20}(Q^2)$
 using twisted mass fermions for $m_\pi\sim 300$~MeV for our  two spatial
lattice sizes of $L=2.1$~fm and $L=2.8$~fm. The lines shown are linear fits
to  the
$Q^2$-dependence. As can be seen,  for both $A_{20}(Q^2)$
and $\tilde{A}_{20}(Q^2)$ one can not ascertain any
volume dependence. For $B_{20}(Q^2)$ and $\tilde{B}_{20}(Q^2)$ the statistical
errors are larger and the linear fits show larger spread with the change in the spatial volume. However, given
the large statistical uncertainties, it is difficult to quantify any volume
dependence.

\subsection{Cut-off effects}

\begin{figure}[h]
{\includegraphics[scale=0.5]{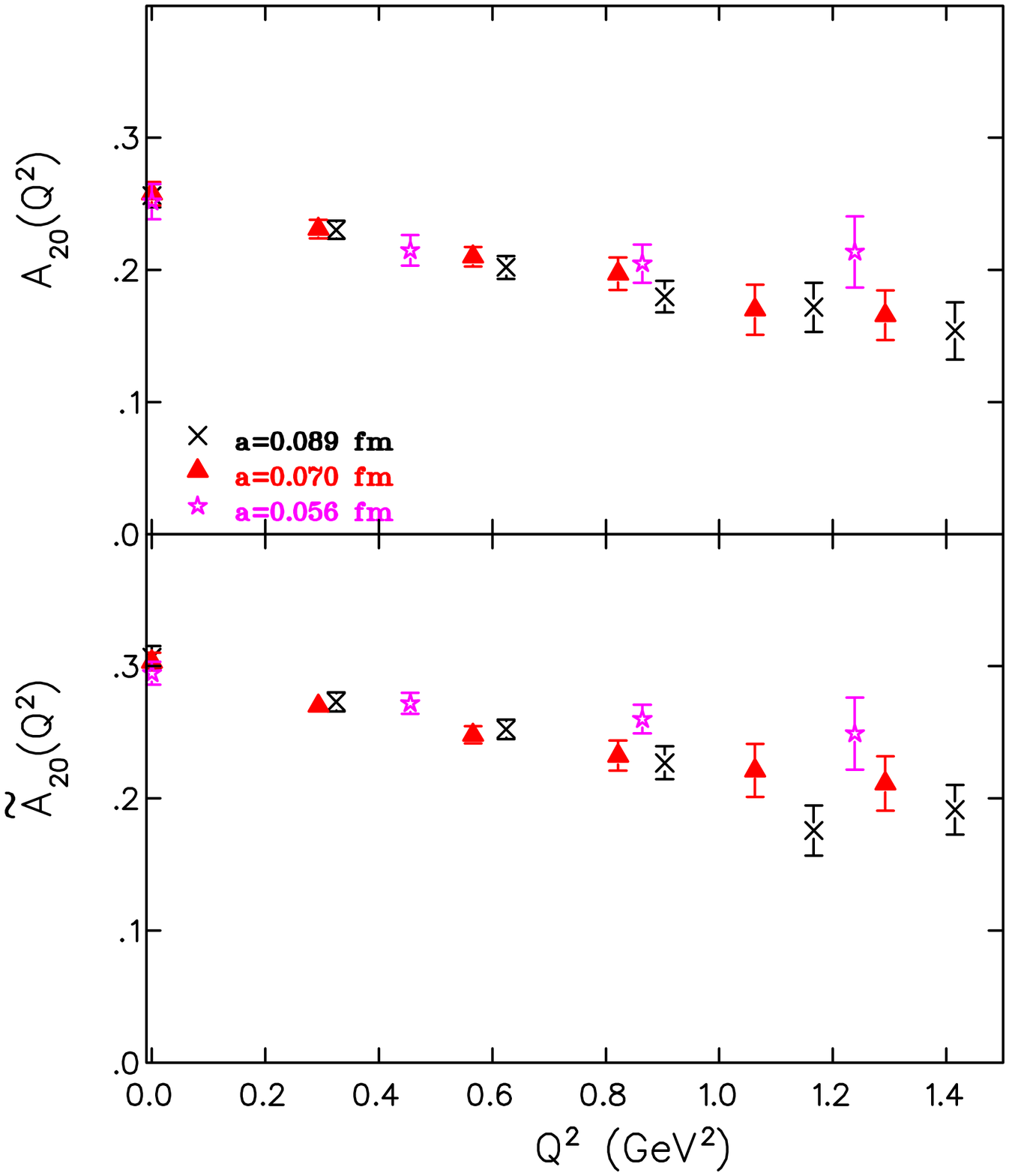}}
{\includegraphics[scale=0.5]{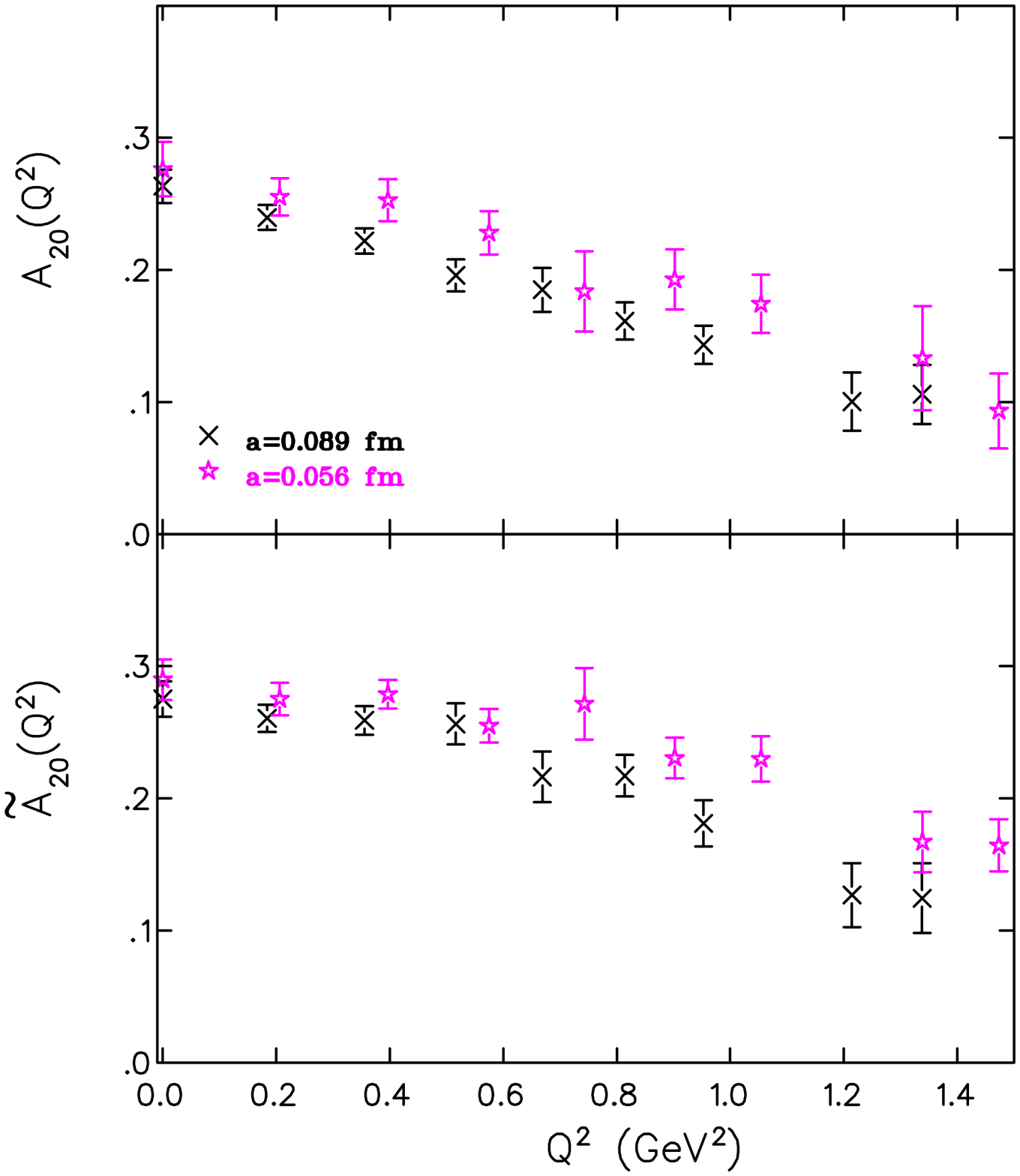}}
\caption{Left panel: The GFFs $A_{20}$ and $\tilde{A}_{20}$ for our three
lattice spacings at $m_\pi\sim 470$~MeV. Right panel: $A_{20}$ and $\tilde{A}_{20}$ for $a=0.089$~fm and $a=0.056$~fm at $m_\pi\sim 260$~MeV. }
\label{fig:a}
\end{figure}

In order to examine cut-off effects we compare in Fig.~\ref{fig:a} our results
obtained  at  the lowest and highest pion mass that we have
considered in this work, namely $m_\pi \sim 260$~MeV and $m_\pi\sim 470$~MeV.
We show results on the quantities $A_{20}(Q^2)$ and $\tilde{A}_{20}(Q^2)$ since
these have smaller statistical errors than $B_{20}(Q^2)$ and $\tilde{B}(Q^2)$.
 For the heavier mass,  where we
have results  at all three lattice spacings, there is no visible dependence
on the lattice spacing especially at low $Q^2$-values.
For the lightest pion mass of $m_\pi=260$~MeV
we have results at the largest and smallest lattice spacings.
The results are in good agreement
although some deviations are seen at larger $Q^2$-values.
Thus, within our current statistical errors,
one may conclude that no significant cut-off effects are
observed.

\subsection{Quark mass dependence}

\begin{figure}[h]
{\includegraphics[scale=0.5]{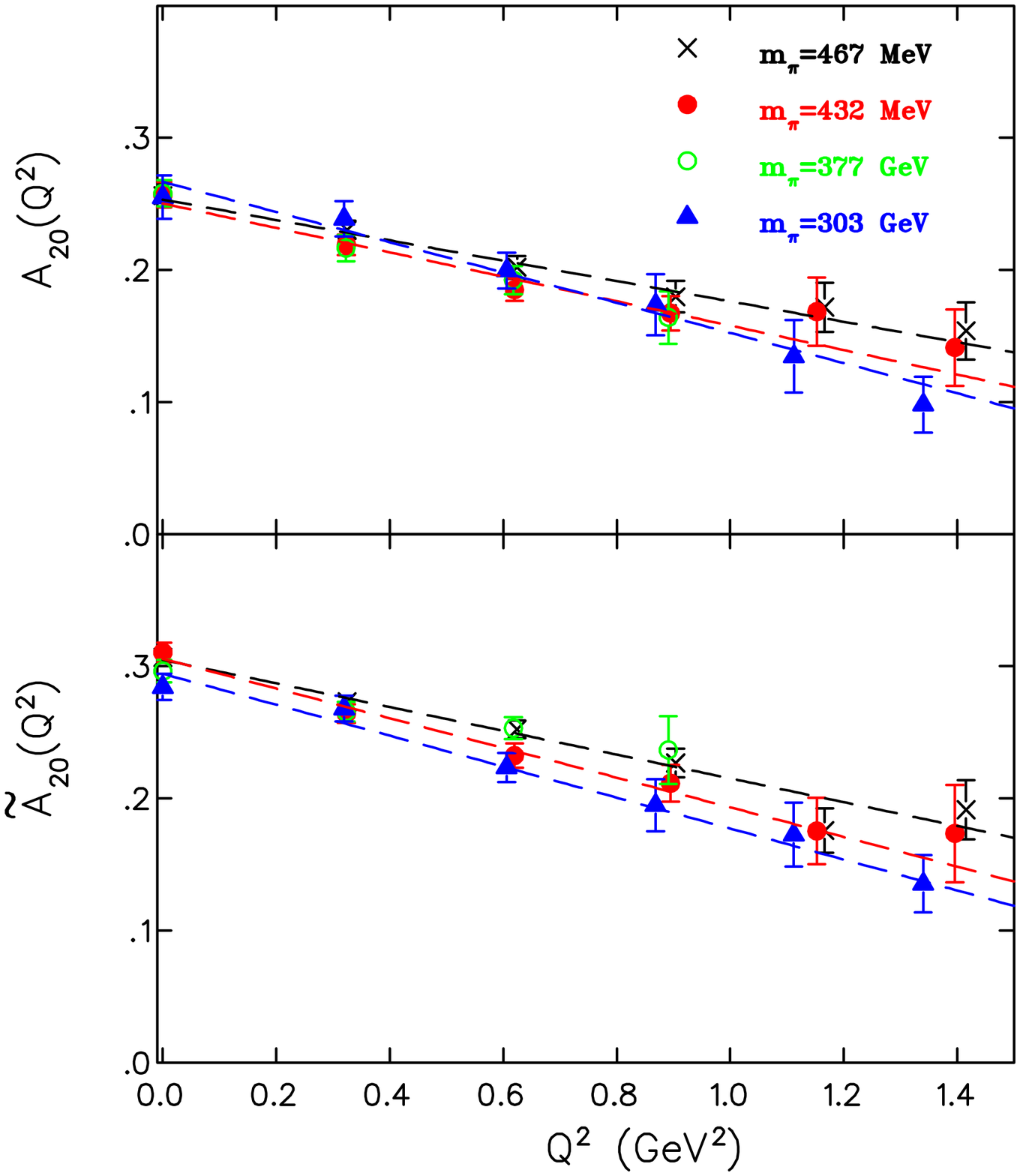}}
\caption{Pion mass dependence of $A_{20}(Q^2)$ and  $\tilde{A}_{20}(Q^2)$
computed at $\beta=3.9$ and using a lattice size of $24^3\times 48$.}
\label{fig:mass}
\end{figure}
The mass dependence of
$A_{20}$ and $\tilde{A}_{20}$ is shown in Fig.~\ref{fig:mass}.
Although the dependence on the mass is weak in the
range of pion  masses spanned, the tendency is for the values of the GFFs to
decrease with decreasing pion mass. There is also a tendency for an increase
in the slope
for both $A_{20}$ and $\tilde{A}_{20}$ as the pion mass decreases.

\subsection{Comparison with other discretization schemes}
\begin{figure}[h]
{\includegraphics[scale=0.49]{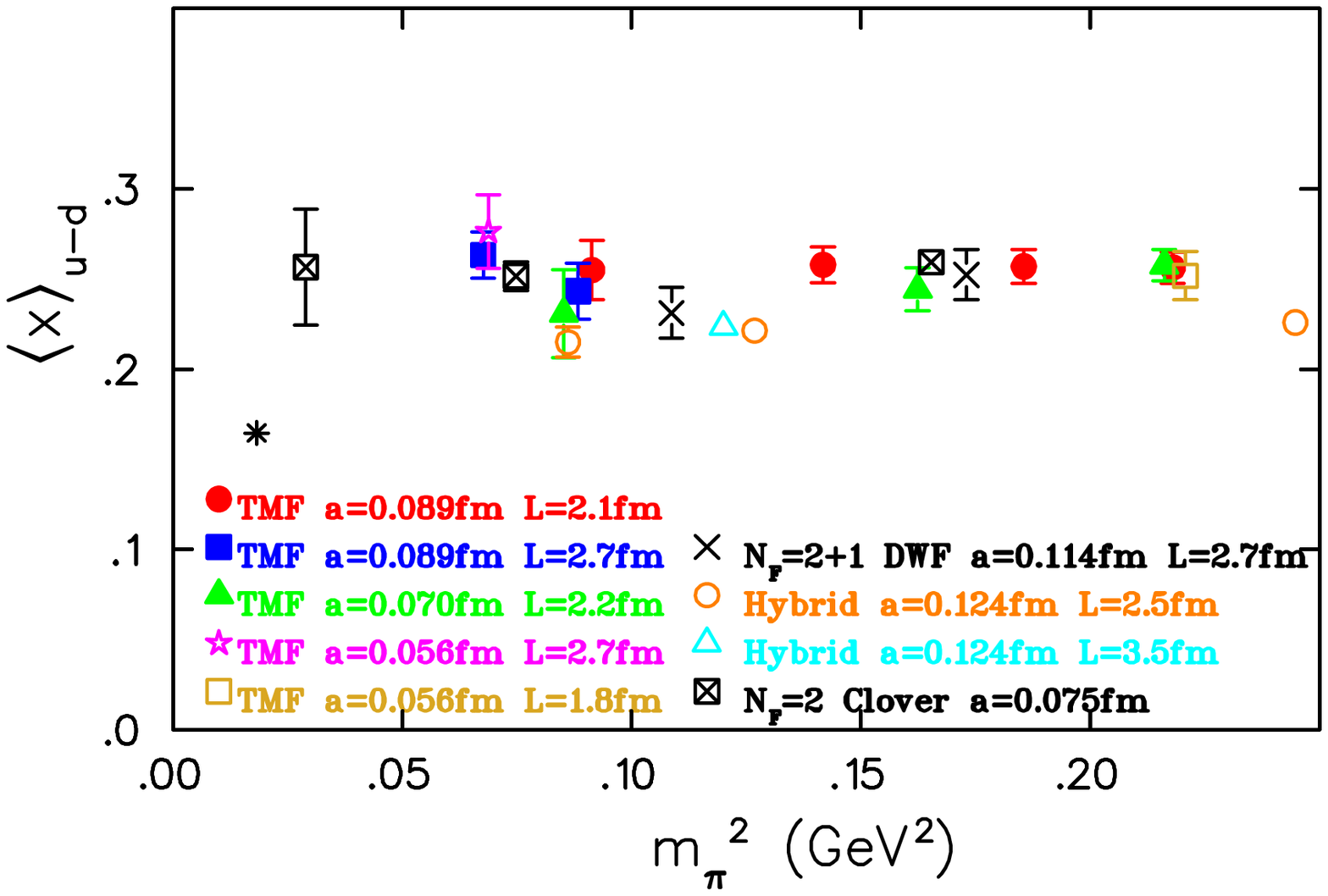}}
{\includegraphics[scale=0.49]{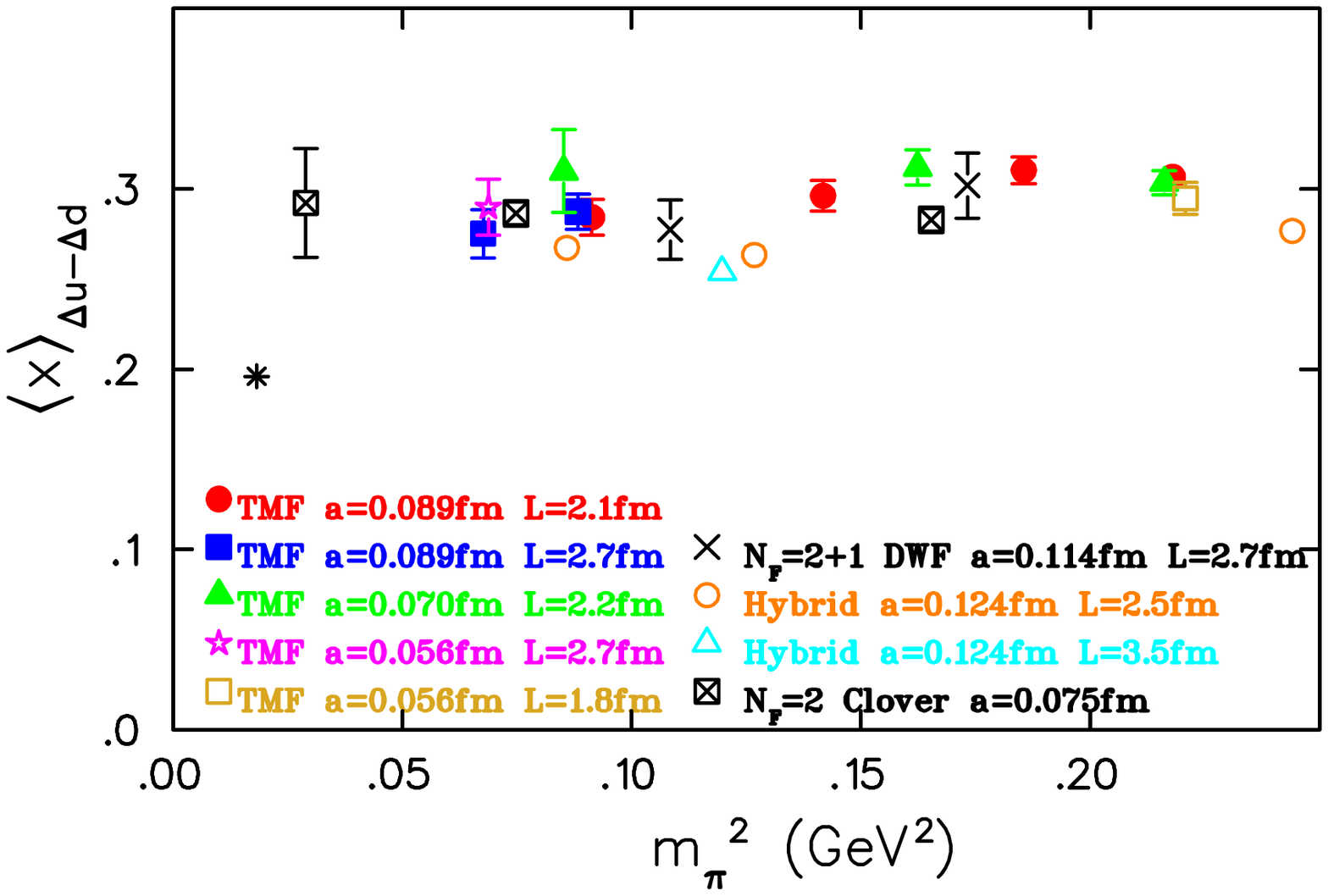}}
\caption{  Lattice data on $\langle x\rangle _{u-d}$ and  $\langle x\rangle _{\Delta u-\Delta d}$ using: (i) $N_F=2$  with $a=0.089$~fm: filled red
circles for $L=2.1$~fm and filled blue squares for $L=2.8$~fm,  $a=0.070$~fm: filled green triangles for $L=2.2$~fm, $a=0.056$~fm:
purple star for $L=2.7$~fm and open yellow square for $L=1.8$~fm;
 (ii)  $N_F=2+1$  DWF~\cite{Aoki:2010xg} crosses for $a=0.114$~fm and $L=2.7$~fm;
(iii) $N_F=2+1$ using DWF for the valence quarks on staggered sea~\cite{Bratt:2010jn} with $a=0.124$~fm: open orange circles for $L=2.5$~fm and open cyan triangle for $L=3.5$~fm;
(iv)  $N_F=2$ clover with
$a=0.075$~fm~\cite{Pleiter:2011gw} (Preliminary results). The physical point, shown by the asterisk, is from Ref.~\cite{Alekhin:2009ni} for the unpolarized and from Ref.~\cite{Airapetian:2009ua,Blumlein:2010rn} for the polarized first moment.}
\label{fig:compare}
\end{figure}

In order to compare lattice data using different discretization schemes
one would have to first extrapolate to the continuum limit.
However, given that the   cut-off effects are small for lattice spacings of
about 0.1~fm,
lattice results for different values of $a$ using a number
of improved discretizations can be directly compared.
In Fig.~\ref{fig:compare} we show
results on the spin-independent and helicity
moments using twisted mass fermions, in the  hybrid approach
obtained by  LHPC~\cite{Bratt:2010jn}, with $N_F=2$ clover fermions
by the QCDSF collaboration~\cite{Pleiter:2011gw} (preliminary results) and with DWF by the
RBC-UKQCD collaborations~\cite{Aoki:2010xg}.
There is good agreement among lattice results. The very accurate results
 obtained  using a hybrid
action of domain wall valence and $N_F=2+1$ staggered fermions~\cite{Bratt:2010jn} tend to be lower
compared to the other
data. One difference between them and the other results presented
 is that they  are perturbatively
renormalized. It was shown in Ref.~\cite{Aoki:2010xg} that
perturbative renormalization can lead to lower values.
The spread in the values of the lattice results is shown to be reduced
by taking
a renormalization free ratio leading
 to a better agreement among lattice data with
$Lm_\pi>4$~\cite{Renner:2010ks}.  In particular, constructing a renormalization
free ratio brought the hybrid data in agreement
with our results using twisted mass fermions and those  using clover fermions by QCDSF.
 Lattice values for $\langle x \rangle_{u-d} =A_{20}(Q^2=0)$ 
although  compatible 
are higher from the phenomenological value $\langle x_{u-d} \rangle \sim 0.16$.
The very recent preliminary result by QCDSF~\cite{Pleiter:2011gw} 
at $m_\pi\sim 170$~MeV
 remains higher than experiment and  highlights the need to understand such deviations. A similar conclusion holds for the helicity moment.

\begin{figure}[h]
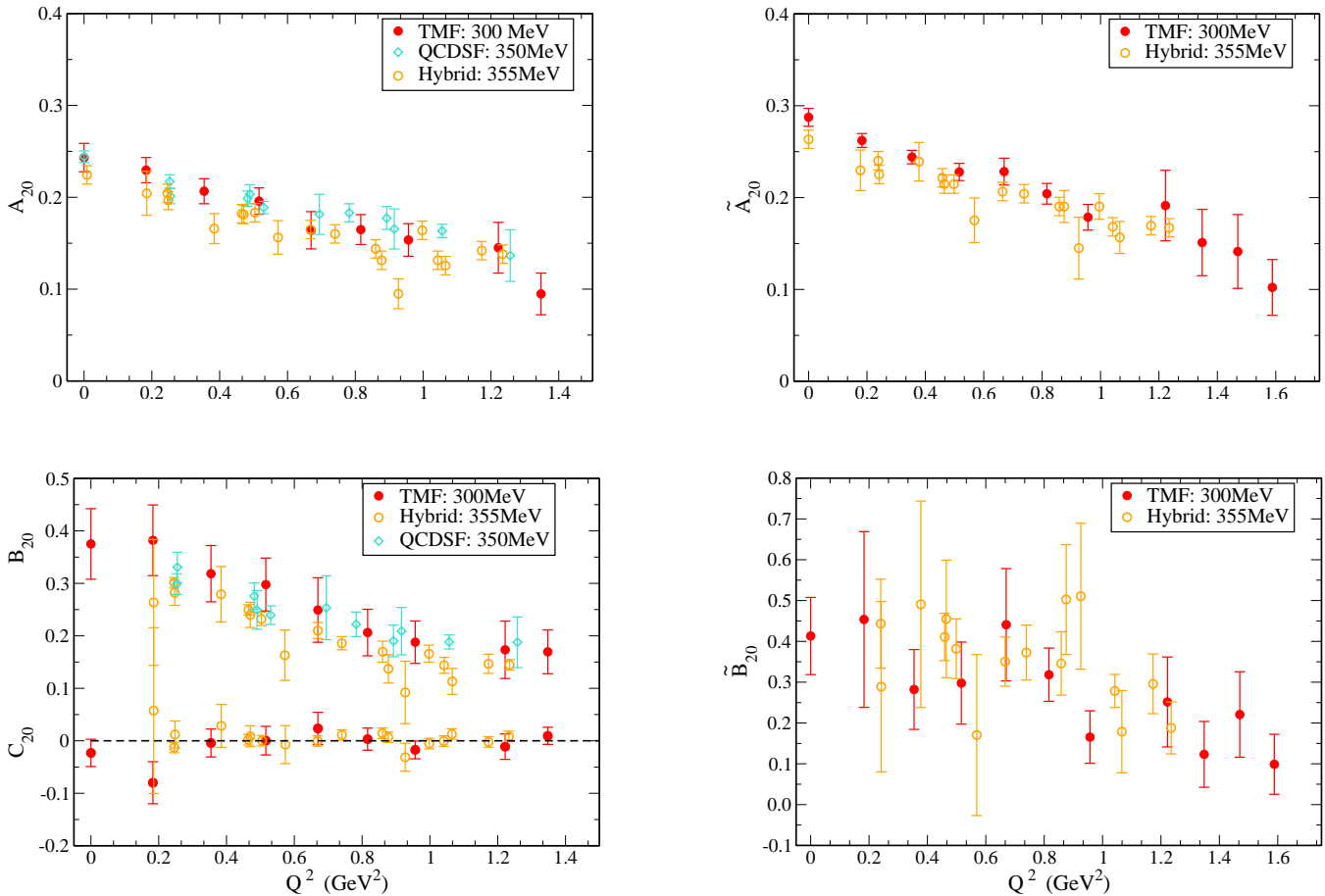

\begin{minipage}{5cm}
{\includegraphics[scale=0.33]{A20_q2_350MeV.eps}}\\\vspace*{0.5cm}
{\includegraphics[scale=0.33]{B20_C20_q2_350MeV.eps}}
\end{minipage}\hfill
\begin{minipage}{8.3cm}
{\includegraphics[scale=0.33]{A20tilde_q2_350MeV.eps}}\\\vspace*{0.5cm}
{\includegraphics[scale=0.33]{B20tilde_q2_350MeV.eps}}
\end{minipage}
\caption{Comparison of twisted mass results for $A_{20}(Q^2)$, $B_{20}(Q^2)$ and $C_{20}(Q^2)$ (left panel) and $\tilde{A}_{20}(Q^2)$ and $\tilde{B}_{20}(Q^2)$ (right panel) at pion mass 300~MeV with those obtained using a hybrid action at $m_\pi=355$~MeV and $N_F=2$
clover fermions at $m_\pi=350$~MeV. }
\label{fig:GFFscompare}
\end{figure}

In Figs.~\ref{fig:GFFscompare} we compare our results for the GFFs  with
pion mass 300~MeV
with those  obtained using a hybrid action by LHPC  and clover fermions
by QCDSF and pion mass of 355~MeV and 350~MeV, respectively.
The results show an overall agreement, with the data of LHPC
somewhat lower than the other two sets. Once more
the fact that both  our results and those of
QCDSF are renormalized non-perturbatively, while those of LHPC are
renormalized perturbatively, might explain this difference.
Moreover, in our determination of the renormalization constants,
we have subtracted ${\cal O}(a^2)$ terms perturbatively
to reduce lattice artifacts~\cite{Alexandrou:2010me}.

\section{Chiral perturbation theory}
In order to make a direct comparison with experiment we need to
extrapolate to the physical point. We 
only perform chiral extrapolation of GFFs at $Q^2=0$.
 We first perform this extrapolation
using our lattice results directly, since, as we have discussed in
the previous section, cut-off effects
are small. In the next section we will perform a continuum extrapolation
and verify that indeed the values we find at the physical point are compatible.

\begin{figure}[h]\vspace*{-0.2cm}
\begin{minipage}{8cm}
\hspace*{-0.5cm}{\includegraphics[width=\linewidth,angle=-90]{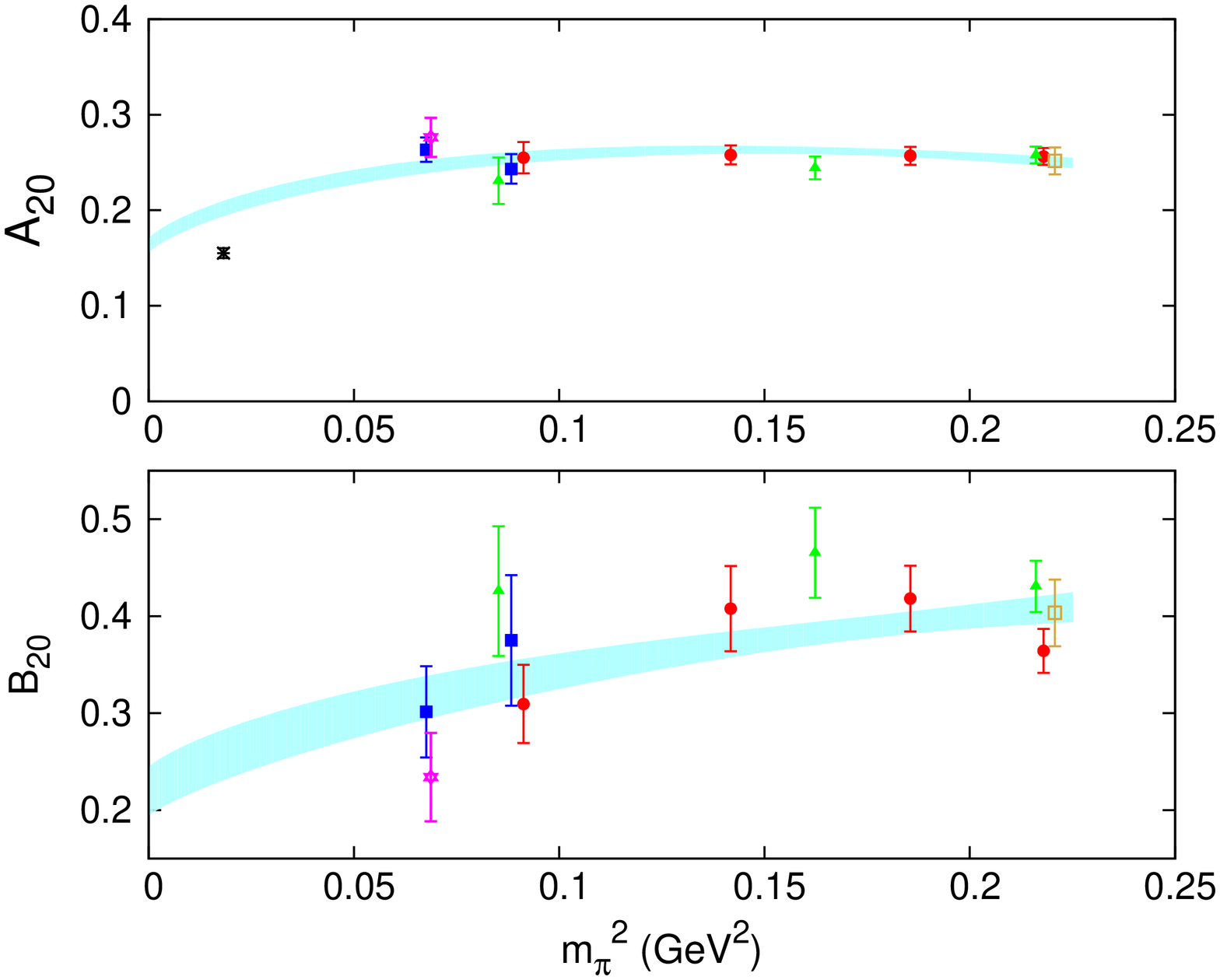}}
\end{minipage}\hfill
\begin{minipage}{8cm}
\hspace*{-0.8cm}{\includegraphics[width=\linewidth,angle=-90]{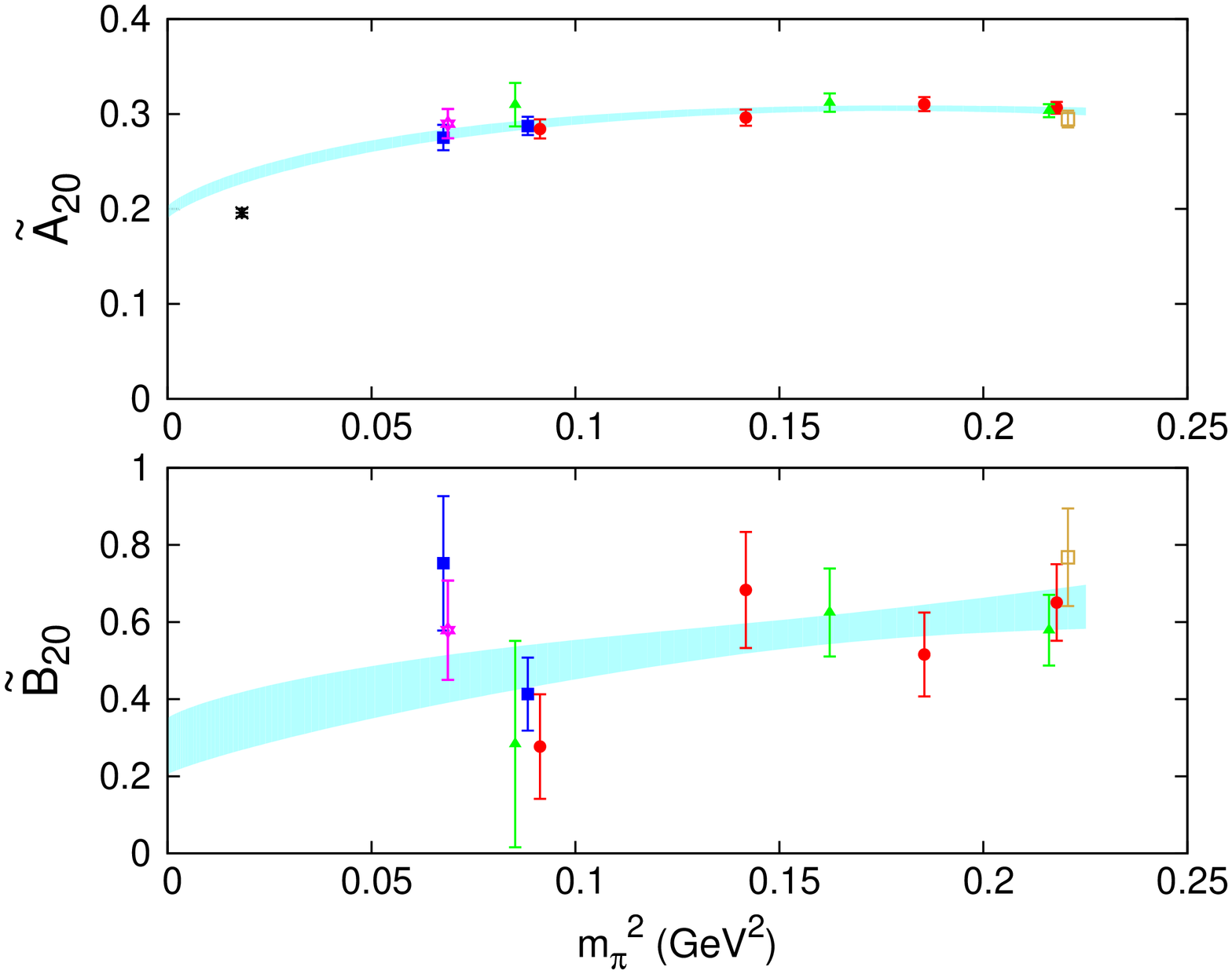}}
\end{minipage}
\caption{Chiral extrapolation using HB$\chi$PT for the isovector unpolarized and polarized first moment of the quark distributions.
The physical point, shown by the asterisk, is from Ref.~\cite{Alekhin:2009ni} for the unpolarized and from Ref.~\cite{Airapetian:2009ua,Blumlein:2010rn} for the polarized first moment.}
\label{fig:GPDs HB}
\end{figure}

\begin{figure}[h]\vspace*{-0.2cm}
\begin{minipage}{8cm}
\hspace*{-0.5cm}{\includegraphics[width=\linewidth,angle=-90]{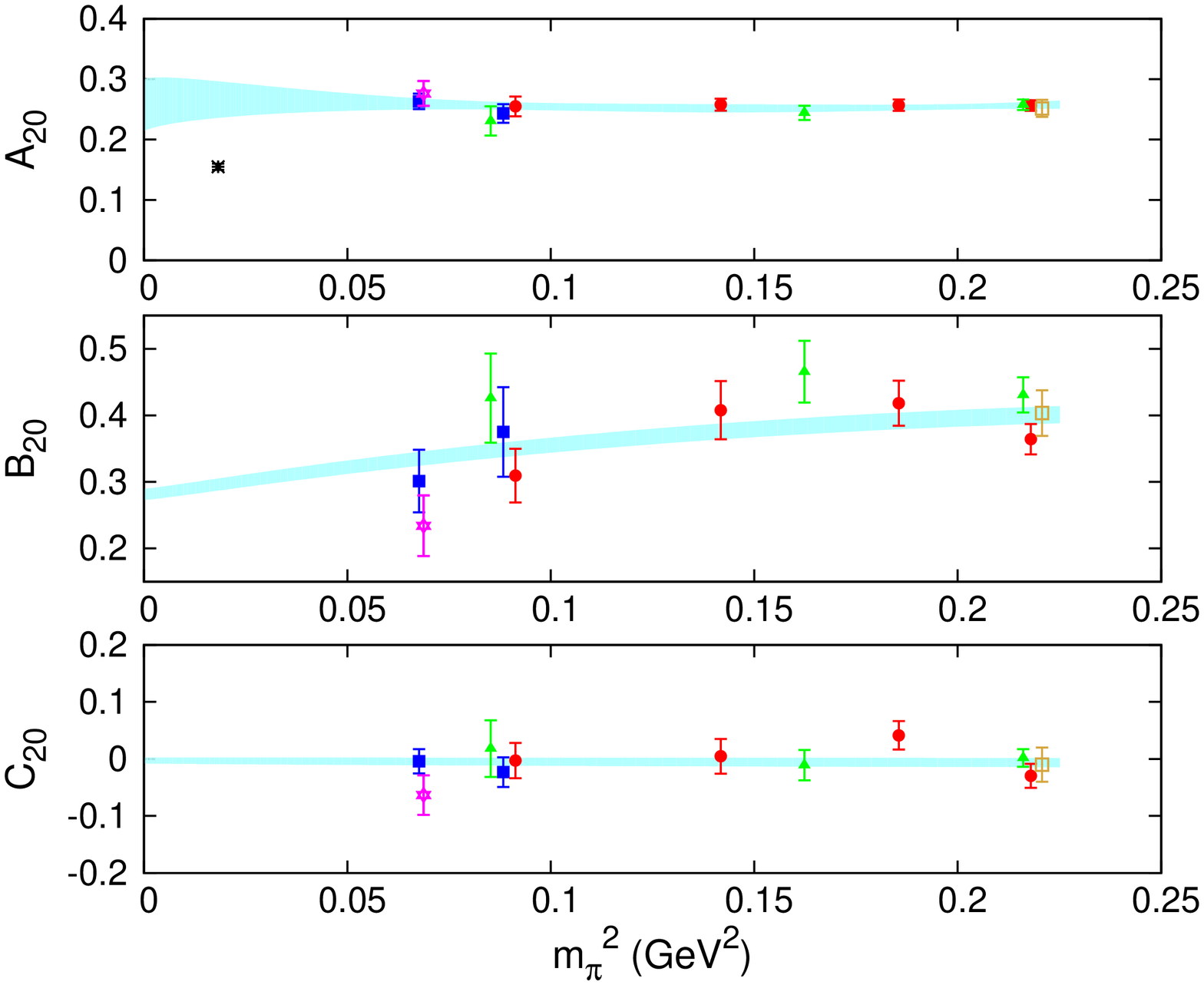}}
\end{minipage}\hfill
\begin{minipage}{8cm}
\hspace*{-0.8cm}{\includegraphics[width=\linewidth,angle=-90]{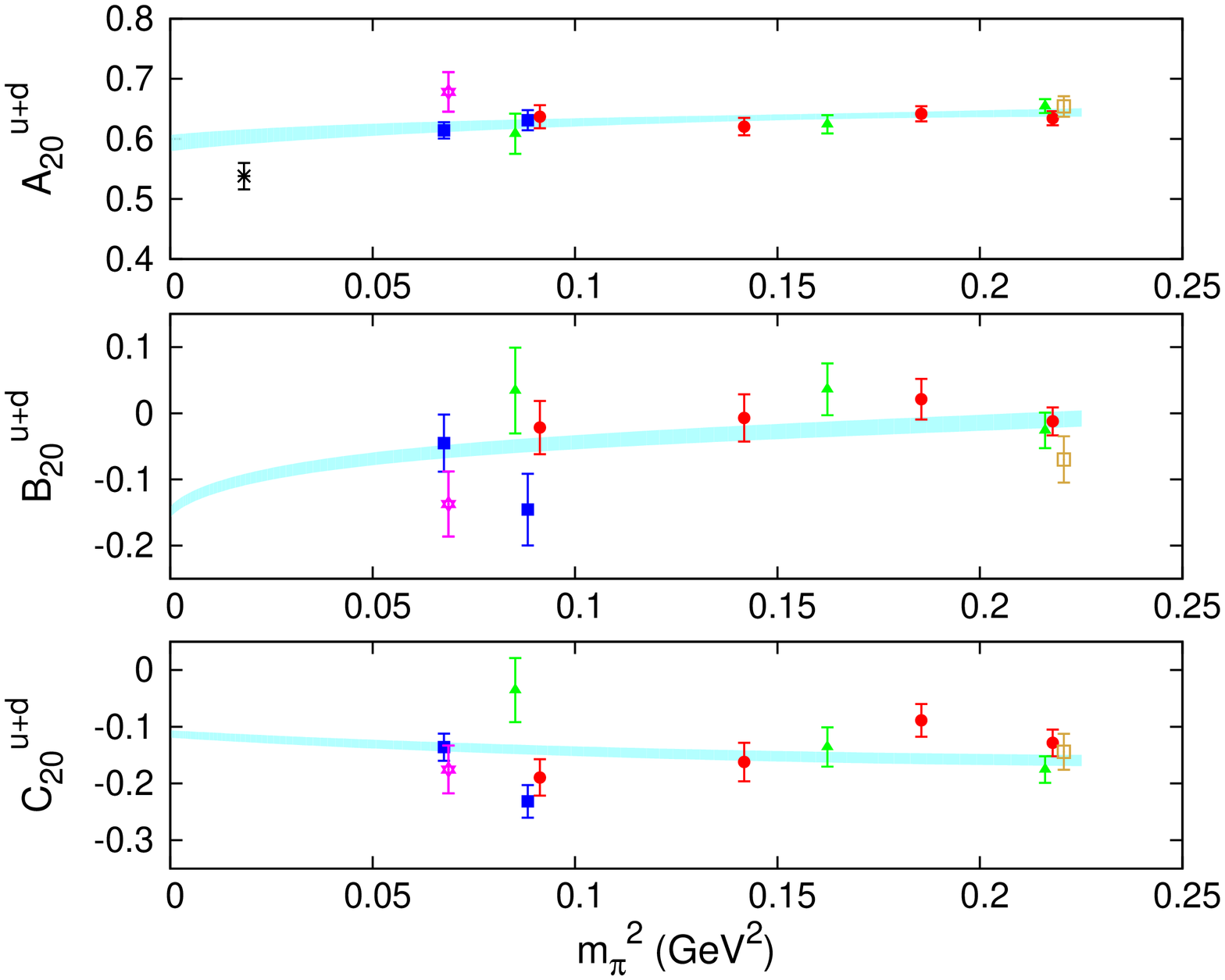}}
\end{minipage}\hfill
\caption{Chiral extrapolation using CB$\chi$PT for the isovector (left) and isoscalar (right)  moments $A_{20}$, $B_{20}$ and $C_{20}$. 
The physical point, shown by the asterisk, is from Ref.~\cite{Alekhin:2009ni} for the isovector and from Refs.~\cite{Pumplin:2002vw,Martin:2002aw} for the isoscalar. }
\label{fig:GPDs CB}
\end{figure}

Within heavy baryon chiral perturbation theory
(HB$\chi$PT)~\cite{Arndt:2001ye,Detmold:2002nf} the expressions for the $m_\pi$-dependence
of $A_{20}$ and $\tilde{A}_{20}$ are given by:
\beq
\langle x \rangle_{u-d} &=& {C}\left[1-\frac{3g_A^2+1}{(4\pi f_\pi)^2}
  m_\pi^2 \ln\frac{m_\pi^2}{\lambda^2} \right] +
\frac{{c_8}(\lambda^2) m_\pi^2}{(4\pi f_\pi)^2}\,, \\
\langle x\rangle_{\Delta u-\Delta d}
&=&{\tilde{C}}\left[1-\frac{2g_A^2+1}{(4\pi f_\pi)^2} m_\pi^2
  \ln\frac{m_\pi^2}{\lambda^2} \right]+ \frac{{\tilde{c}_8}(\lambda^2)
  m_\pi^2}{(4\pi f_\pi)^2},
\label{HBchPT}
\eeq
where we take $\lambda^2=1$~GeV$^2$.  The best fit is  shown in
Fig.~\ref{fig:GPDs HB}, where the width of the band
is computed through a super-jackknife analysis~\cite{Bratt:2010jn}.
As can be seen, the fits yield a value higher than experiment for both observables.
LHPC carried out a combined chiral fit  using
$\Op(p^2)$ covariant baryon chiral perturbation theory  (CB$\chi$PT)~\cite{Dorati:2007bk} to $A_{20}$, $B_{20}$ and $C_{20}$.
The mass of the nucleon at the chiral limit is used   as input to the
fits. They
obtained a value for $A_{20}$ in agreement with experiment~\cite{Bratt:2010jn}.
In order to compare with their analysis we also perform a combined
fit to  $A_{20}(0)$, $B_{20}(0)$ and $C_{20}(0)$
within  CB$\chi$PT~\cite{Dorati:2007bk}. 
The CB$\chi$PT fits are shown by the bands in Fig.~\ref{fig:GPDs CB}.
As can be seen, they also provide a good description to the lattice data 
but, in the case of $A_{20}$, CB$\chi$PT leads to
an even higher value at the physical point. Therefore, the discrepancy between our lattice results
and the experimental value is not resolved.
In Appendix B we collect the formulae
used for the chiral extrapolations.
The actual
renormalized lattice data  are tabulated in the tables of Appendix C
for the isovector GFFs.


\section{Continuum extrapolation}
In order to study the continuum extrapolation we use the simulations at our three lattice
spacings at the smallest and largest pion mass.
We first interpolate the GFFs
 at the three values of $\beta$  to a given value of the pseudoscalar
mass
in units of $r_0$.
  We take
as reference pion masses the ones computed on the finest lattice and
interpolate results at the other two $\beta$-values to these two reference masses.

As already mentioned, 
 $<x>_{u-d}=A_{20}(0)$ and  $\langle x\rangle_{\Delta
   u-\Delta d}=\tilde{A}_{20}(0)$ are calculated directly at $Q^2=0$
 requiring no fits.  We therefore
choose these quantities to examine their dependence on the lattice
 spacing since this choice avoids any systematic errors due to the
extrapolation to $Q^2=0$, which would require the adaptation of an Ansatz
for the $Q^2$-dependence.


\begin{figure}
{\includegraphics[scale=0.32]{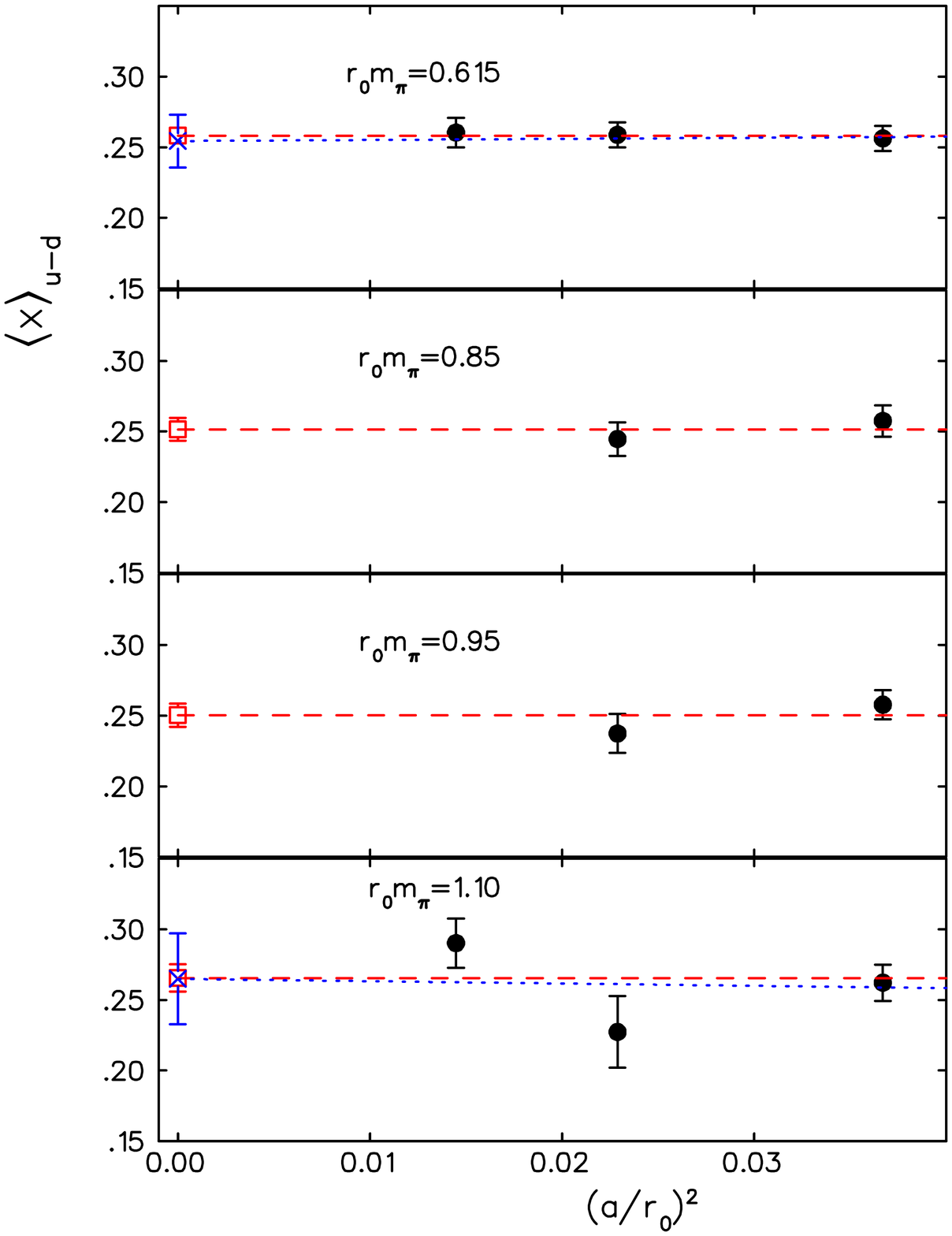}}
{\includegraphics[scale=0.32]{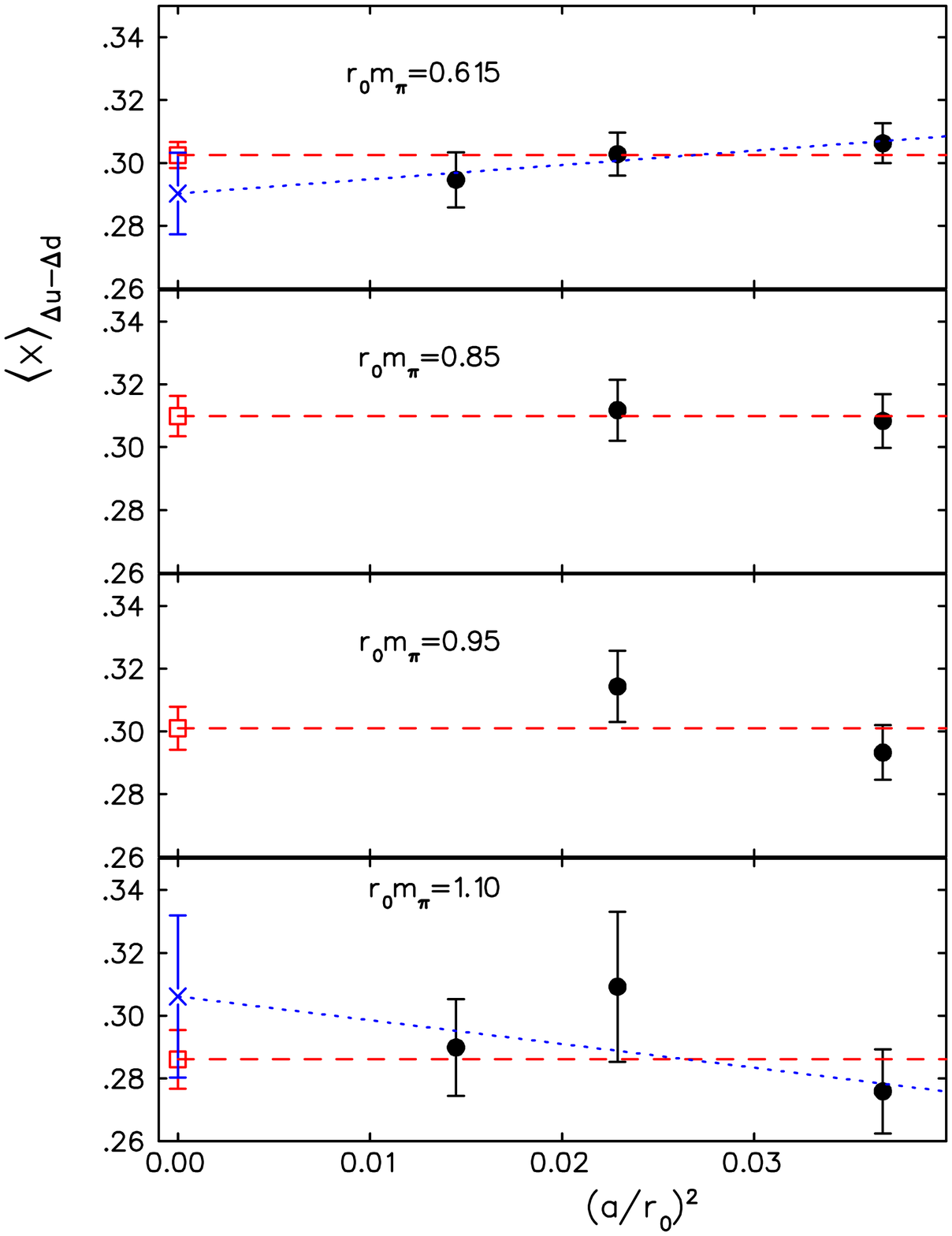}}
{\includegraphics[scale=0.32]{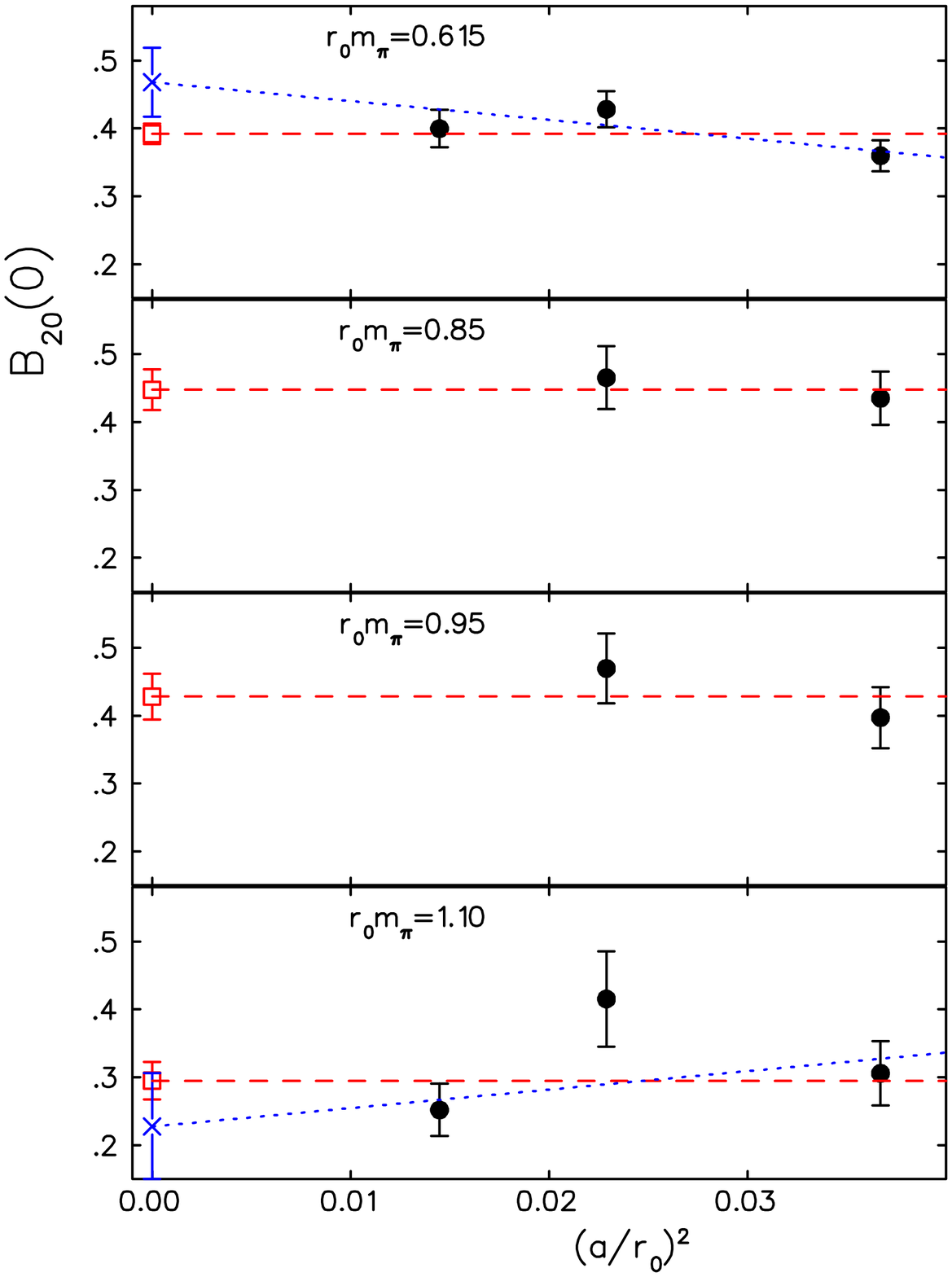}}
\caption{$\langle x \rangle_{u-d}$,  $\langle x \rangle_{\Delta
    u-\Delta d}$ and $B_{20}(0)$ as a function of $(a/r_0)^2$. The red
line is the result of fitting to a constant; the blue one is a linear
fit. }
\label{fig:cont}
\end{figure}

\begin{figure}
{\includegraphics[scale=0.49]{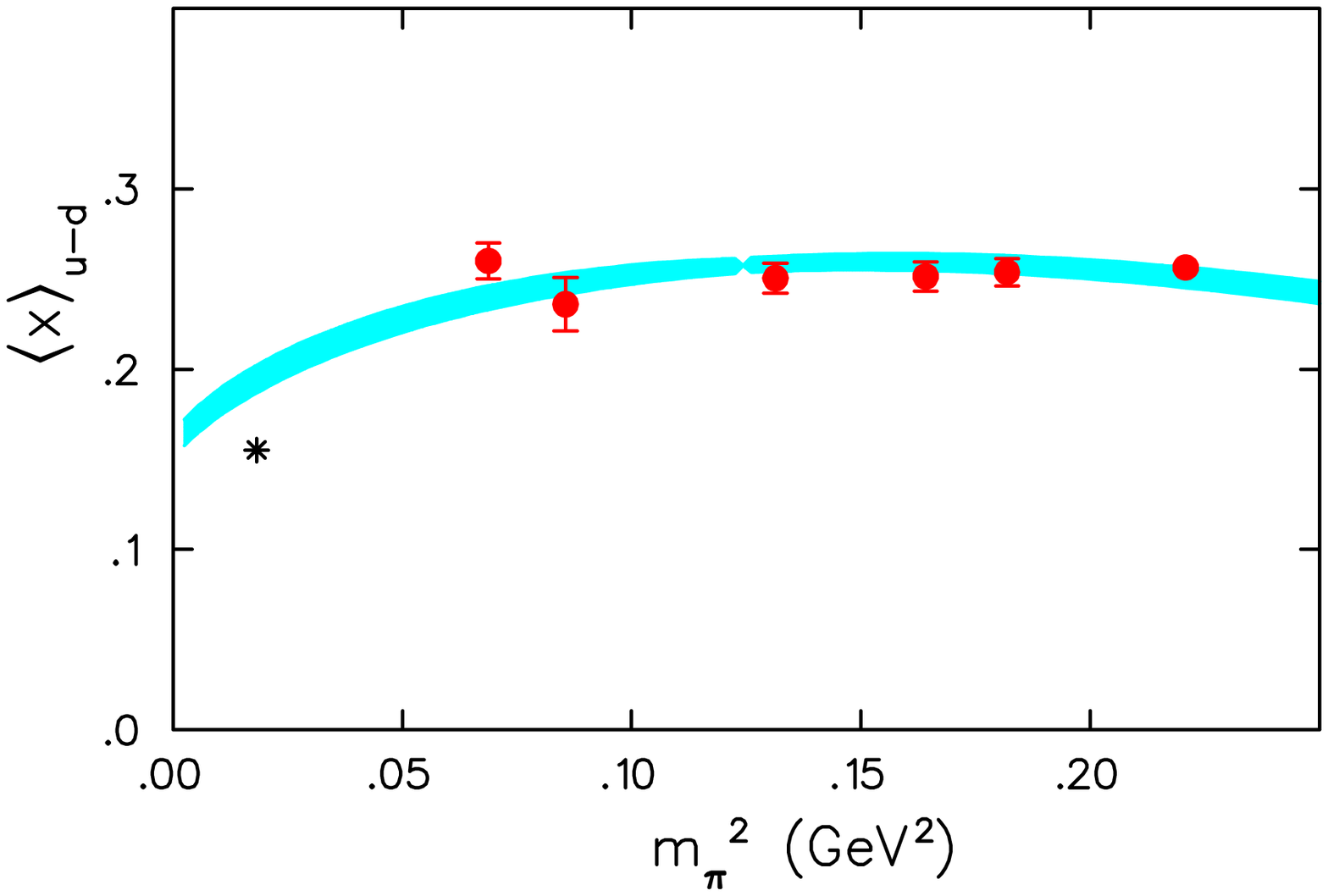}}
{\includegraphics[scale=0.49]{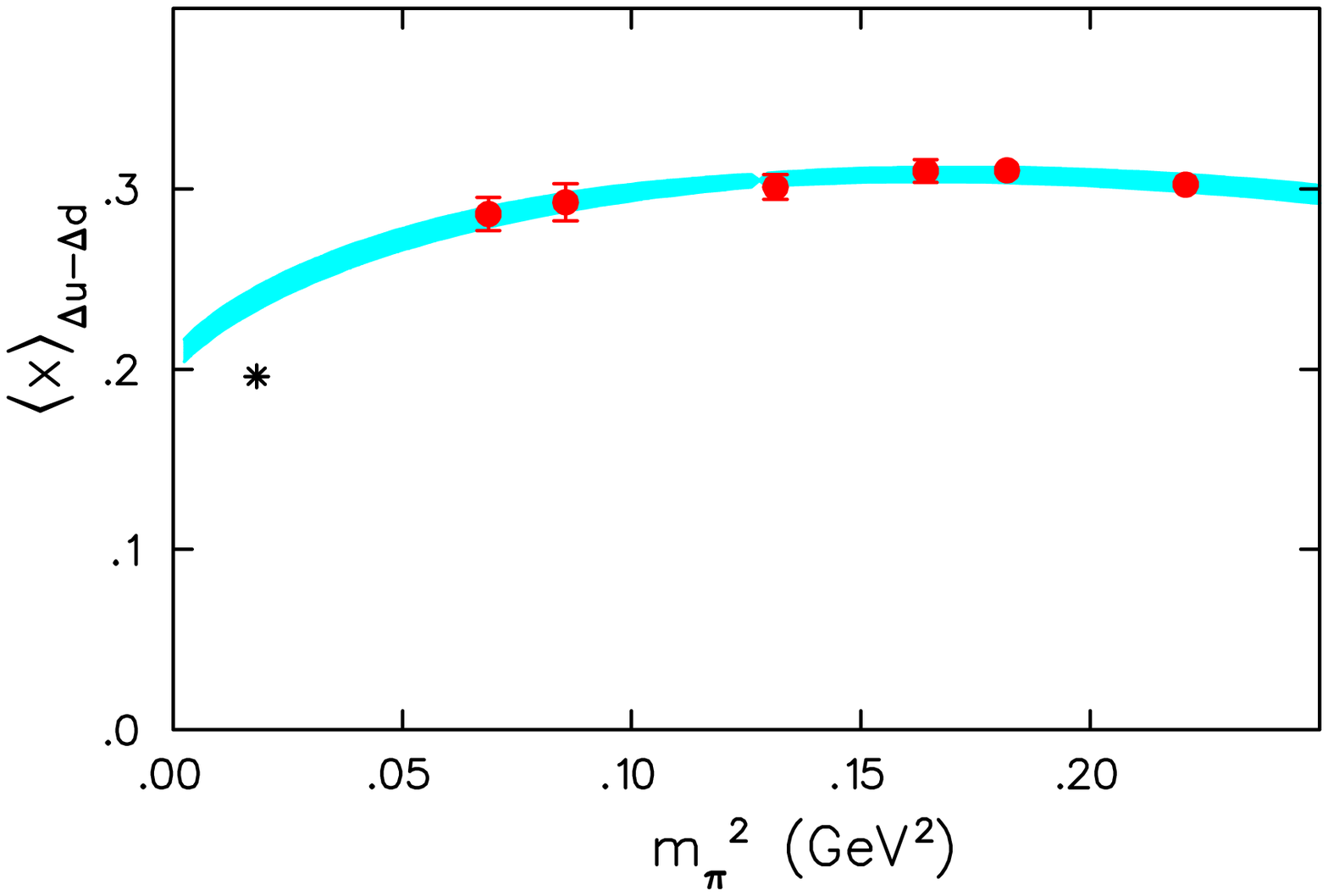}}
\caption{$\langle x \rangle_{u-d}$ (left) and $\langle x \rangle_{\Delta
    u-\Delta d}$ (right)  extrapolated to $a=0$ as a function of $m_\pi$.
The blue band is the chiral fit using HB$\chi$PT.The physical point, shown by the asterisk, is from Ref.~\cite{Alekhin:2009ni} for the unpolarized and from Refs.~\cite{Airapetian:2009ua,Blumlein:2010rn} for the polarized first moment.}
\label{fig:cont fits}
\end{figure}

Having determined the values at a given reference pion mass, we perform a fit to
these data using the form $y(a)=y(0)+c(a/r_0)^2$. The resulting
fits are shown in Fig.~\ref{fig:cont}. Setting $c=0$ we obtained
the constant line also shown in the figure. As can be seen, for both large
and small pion masses allowing a non-zero
slope yields a value in the
continuum limit that is in agreement with that obtained using a constant fit.
This analysis shows  that finite $a$ effects are small for both large
and small pion masses and
extrapolation to the continuum limit using a constant fit is acceptable.
For the intermediate pion masses we therefore
obtained the values in the continuum by fitting our data at $\beta=3.9$ and $\beta=4.05$ to a constant.
For comparison we also perform a similar analysis for $B_{20}(0)$ which 
requires fitting the $Q^2$-dependence.
The qualitative behavior is similar to the one observed for  $A_{20}(0)$ and
$B_{20}(0)$.

Having results at the continuum limit we perform a chiral fit using
HB$\chi$PT. The resulting curves are shown in Fig.~\ref{fig:cont fits} and
still produce a value at the physical point that is higher than the experimental
value. In fact, the value obtained at the physical point
 for both vector and axial-vector moments
is in agreement to the one extracted using the raw lattice data. This
provides an {\it a posteriori} justification of using continuum chiral
perturbation theory directly on the lattice data obtained at our three lattice
spacings  to perform the extrapolation to the physical point in the previous section.

\section{Proton Spin}
In order to extract information on the spin content of the nucleon
one needs to evaluate the isoscalar moments $A_{20}^{u+d}$ and $B_{20}^{u+d}$ since the total spin of a quark in the nucleon is given by
\be J^q=\frac{1}{2}\left ( A_{20}^q (0)+
B_{20}^q(0) \right).
\ee
The total spin can be further decomposed into its orbital angular momentum $L^q$
and its spin component $\Delta\Sigma^q$ as
\be
J^q=\frac{1}{2}\Delta\Sigma^q +L^q
\ee
The spin carried by
the u- and d- quarks is
determined using  $\Delta\Sigma^{u+d}=\tilde{A}_{10}^{u+d}$.
In order to evaluate the isoscalar quantities
one would need the disconnected contributions.
These are notoriously difficult to calculate and they are neglected
in most current evaluations of GFFs. Under the assumption that these
are small we may extract the information on the nucleon spin.



\begin{figure}[h]\vspace*{-0.2cm}
\begin{minipage}{6cm}
{\includegraphics[width=1.1\linewidth,angle=-90]{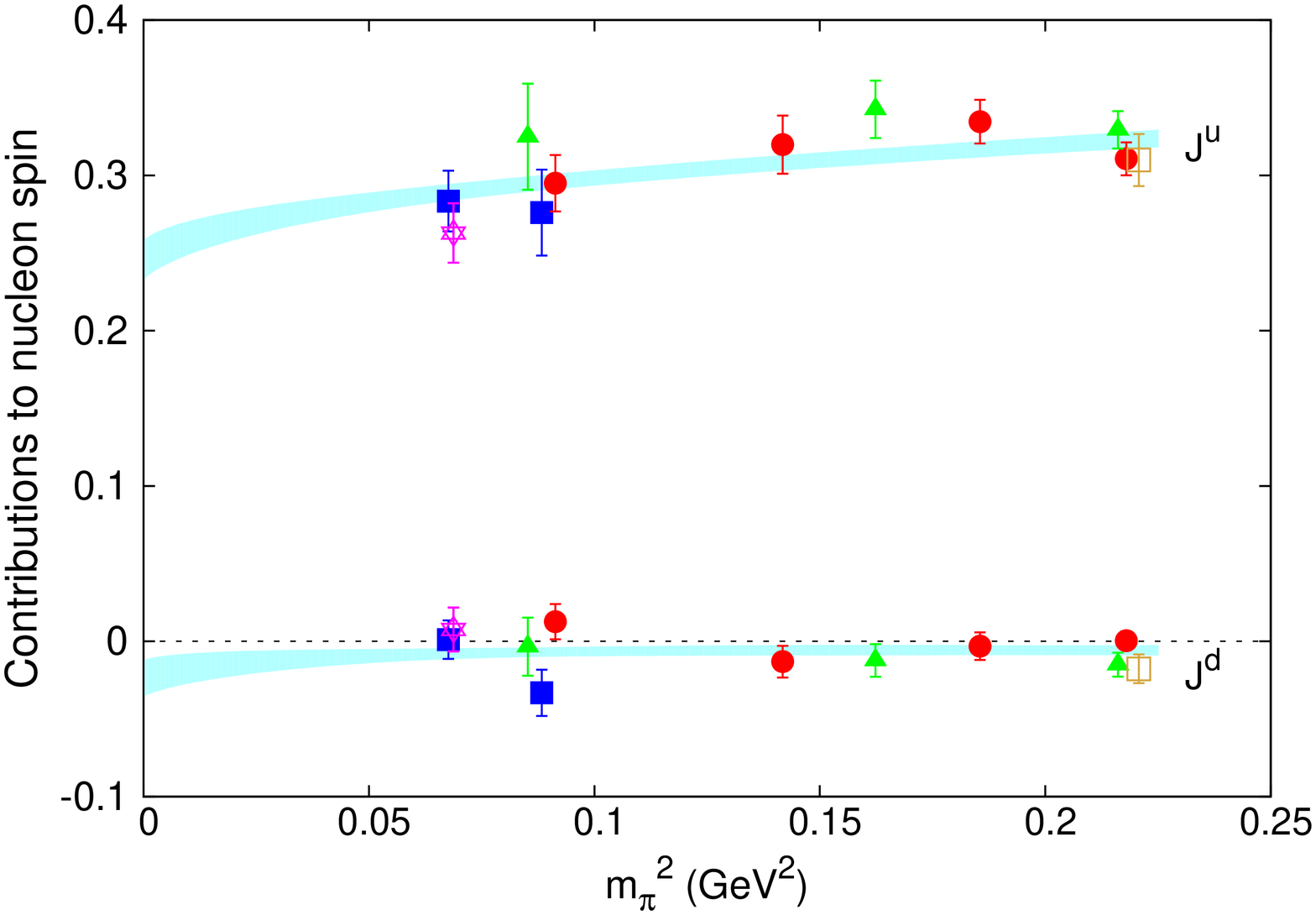}}
\end{minipage}\hfill
\begin{minipage}{6cm}
\hspace*{-2cm}{\includegraphics[width=1.1\linewidth,angle=-90]{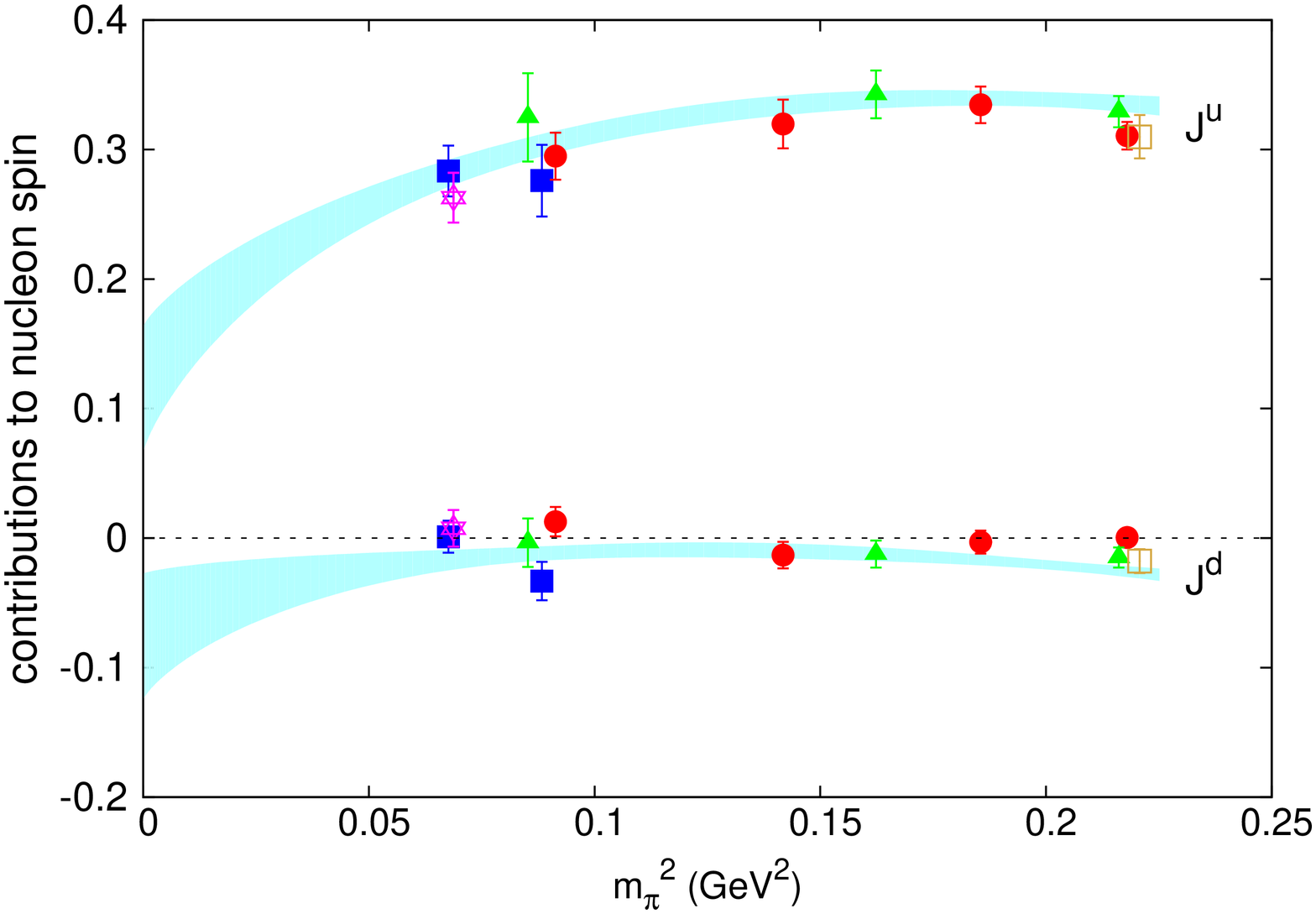}}
\end{minipage}\hfill
\caption{Chiral extrapolation using CB$\chi$PT (left) and HB$\chi$PT (right) for the total spin carried by the u-and d- quarks.}
\label{fig:J CB HB}
\end{figure}

\begin{figure}[h]\vspace*{-0.2cm}
\begin{minipage}{6cm}
{\includegraphics[width=\linewidth,angle=-90]{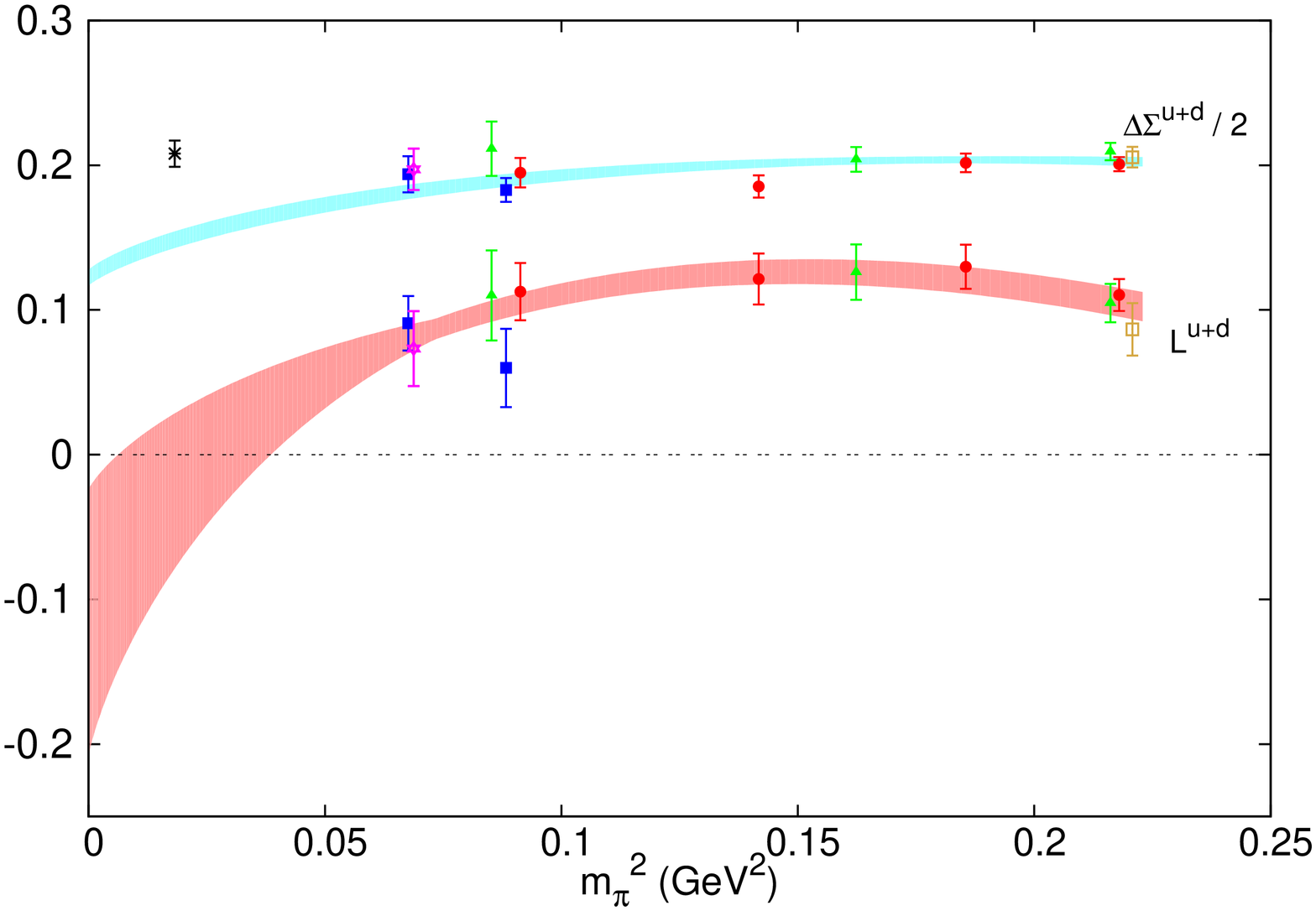}}
\end{minipage}\hfill
\begin{minipage}{6cm}
\hspace*{-2cm}{\includegraphics[width=\linewidth,angle=-90]{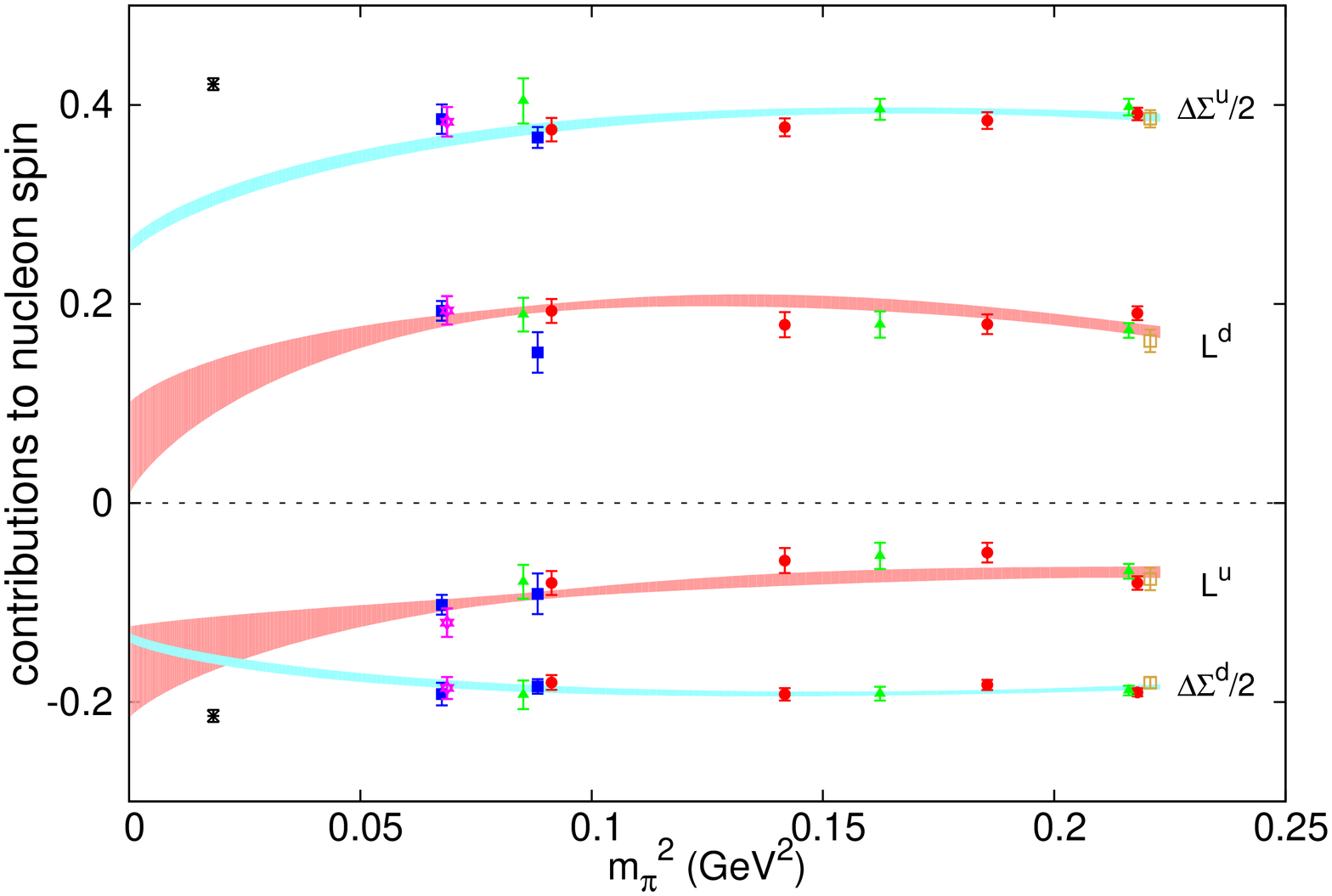}}
\end{minipage}\hfill
\caption{Chiral extrapolation using HB$\chi$PT  for the angular momentum and
 spin carried by the u-and d- quarks. The physical points, shown by the asterisks are from the HERMES 2007 analysis~\cite{Airapetian:2006vy}. }
\label{fig:L and S}
\end{figure}

\begin{table}
\begin{center}
\begin{tabular}{c||c||c||c}
\hline
\hline
                		&   CB$\chi$PT   &  HB$\chi$PT      & experiment \\
\hline                
 $J^{u-d}$				&            &   0.236(14)  &            \\
 $J^{u+d}$      	    &            &   0.143(56)  &            \\
 $J^u    $ 	     	    &  0.266(9)  &   0.189(29)  &            \\
 $J^d    $      	    & -0.015(8)  &  -0.047(28)  &            \\
 $\Delta\Sigma^{u-d}/2$ &            &   0.462(11)  &            \\
 $\Delta\Sigma^{u+d}/2$ &            &   0.148(5)   &  0.208(9)  \\
 $\Delta\Sigma^u/2$     &            &   0.305(7)   &  0.421(6)  \\
 $\Delta\Sigma^d/2$     &            &  -0.157(5)   & -0.214(6)  \\
 $L^{u-d}$              &            &  -0.258(5)   &            \\
 $L^{u+d}$              &            &  -0.025(53)  &            \\
 $L^u$                  &            &  -0.141(30)  &            \\
 $L^d$                  &            &   0.116(27)  &            \\
\hline\hline
\end{tabular}
\caption{Values of nucleon spin observables at the physical point using
CB$\chi$PT and HB$\chi$PT and from experiment~\cite{Airapetian:2006vy}.} 
\label{tab:spin parameters}
\end{center}
\end{table}

In Fig.~\ref{fig:GPDs CB}  we show our results for 
the isoscalar $A_{20}(0)^{u+d}$, $B_{20}(0)^{u+d}$
and $C_{20}(0)^{u+d}$. Since as shown in the previous section,
 cut-off effects  are
  small, we here perform a chiral extrapolation  
 directly on the lattice data. Having both isoscalar and isovector quantities we can extract the spin $J^u$ and $J^d$ carried by the u- and d- quarks.
The results are shown in Fig.~\ref{fig:J CB HB}. We show the extrapolation
using both HB$\chi$PT and CB$\chi$PT both of which have the same
qualitative behavior.  As can be seen, the
contribution to the spin from the d-quark is much smaller than that of the
u-quarks. 
These results are in qualitative agreement with the recent results
obtained using a hybrid action~\cite{Bratt:2010jn}.
In Fig.~\ref{fig:L and S} we show separately the orbital angular momentum
and spin carried by the u- and d- quarks. Both our results and those
of LHPC~\cite{Bratt:2010jn} are in qualitative agreement as far as the spin
is concerned.
 For the orbital angular
momentum
 we obtain higher values for both d- and u-quark (less negative). Thus
we obtain a total positive $L^{u+d}$ compared to a small negative value in the
case of LHPC. After chiral extrapolation, the value obtained at the 
physical point is consistent with zero  in agreement with the
value determined by LHPC. For the spin contribution
$\Delta \Sigma^{u+d}$ our value is lower at the physical
point as compared to  that obtained by LHPC.
We summarize the values for the total spin, orbital angular
momentum and spin in the proton at the physical point in Table~\ref{tab:spin parameters}.

\section{Conclusions}
We have performed  an analysis  on the
generalized form factors, $A_{20}(Q^2),\,B_{20}(Q^2),\,C_{20}(Q^2),\,\tilde A_{20}(Q^2),\,\tilde B_{20}(Q^2)$, extracted from the
nucleon matrix elements of  the one-derivative vector and axial-vector
operators
using two degenerate flavors of twisted mass fermions.
Our results are non-perturbatively renormalized and they are presented
in the $\overline{\rm MS}$ scheme at a scale of 2 GeV.
To investigate  volume and cut-off effects we have used the isovector combinations, which can be calculated without the necessity to  evaluate 
disconnected contributions. Our main conclusion regarding cut-off effects is that they are small within the current accuracy of about 5-10\% and for
lattice spacings smaller than 0.1~fm. Similarly, no systematic
volume effects are seen. Given the small cut-off effects one can
compare lattice results directly using different discretization schemes.
The comparison of the results using $N_F=2$ twisted mass fermions
with the results obtained using $N_F=2$ clover fermions by the
QCDSF~\cite{Pleiter:2011gw} shows agreement. Both the results of this work
as well as those by QCDSF are non-perturbatively renormalized.
We also compared our results with $N_F=2+1$ domain wall fermions~\cite{Aoki:2010xg}. Again
there is agreement without any indication of any systematic effect from
including a dynamical strange quark.
Our results at three values of the lattice spacing allow for a continuum extrapolation. By interpolating results to a reference mass in units of $r_0$ and performing a linear extrapolation in $a^2$ it was shown that the values obtained are consistent with those obtained with a constant extrapolation. This has been verified for both the heaviest and lightest masses used in this work.
 Furthermore, if one
performs chiral fits to the extrapolated continuum results one finds
a value at the physical point consistent with the one obtained
 using directly the lattice
data at finite lattice spacing. This provides a consistency check that
cut-off effects for a lattice spacing less than 0.1~fm
are smaller than our current statistical errors.

Having established that both volume and  cut-off effects are small for the isovector quantities for which only connected contributions are needed, we analyse
the corresponding isoscalar quantities using directly our lattice data. Of particular interest here is the spin content of the nucleon. The disconnected
contributions to the isoscalar quantities are not included.
We find that the spin carried by  the d-quark is almost zero whereas the
u-quarks carry about 50\% of the nucleon's spin. This result is consistent
with other lattice calculations~\cite{Bratt:2010jn}.

For the chiral extrapolations of these quantities we use HB$\chi$PT and CB$\chi$PT theory. In both cases, our results on the momentum
fraction and helicity moment   at the physical point are higher than their
experimental value. Such discrepancies are also observed
in the case of the nucleon axial charge and they  need to
be further investigated.

\section*{Acknowledgments}
We would like to thank all members of ETMC for a
very constructive and enjoyable collaboration and for the many fruitful
discussions that took place during the development of this work.

Numerical calculations have  used HPC resources from GENCI (IDRIS and CINES) Grant 2009-052271  and
CC-IN2P3 as well as  from the
John von Neumann-Institute for Computing on the Juropa and  Jugene
systems at the research center
in J\"ulich. We thank the staff members for their kind and sustained support.
M.Papinutto acknowledges financial support by the Marie Curie European
Reintegration Grant of the 7th European Community Framework
Programme under contract number PERG05-GA-2009-249309.
This work is supported in part by  the DFG
Sonder\-for\-schungs\-be\-reich/ Trans\-regio SFB/TR9 and by funding received from the
 Cyprus Research Promotion Foundation under contracts EPYAN/0506/08,
KY-$\Gamma$/0907/11/ and TECHNOLOGY/$\Theta$E$\Pi$I$\Sigma$/0308(BE)/17.

\bibliography{GPDs_ref}

\newpage

{\bf{Appendix A: Expressions for the extraction of  GFFs from lattice measurements}}

\vspace*{0.5cm}

We collect here the expressions relating the plateau values to the GFFs
$A_{20}$, $B_{20}$, $C_{20}$ and $\tilde{A_{20}}$, $\tilde B_{20}$. 
The index $V$ ($A$) refers to the vector (axial-vector) one-derivative operator. 
All relations are given in Euclidean space.

{\bf{\underline{Vector:}}}
\begin{eqnarray}
   \Pi_V^{00}(\Gamma^0, \vec q) &=& A_{20}\,C\,\left(-\frac{3\,E_N}{8} - \frac{E_N^2}{4\,m_N} -
     \,\frac{m_N}{8} \right) +
 B_{20}\,C\,\left( -\,\frac{E_N }{8} +
      \,\frac{E_N^3}{8\,m_N^2} + \frac{E_N^2}{16\,m_N} - \frac{m_N}{16}
      \right)\nonumber\\ [2.15ex]
&+&   C_{20}\,C\,\left(\,\frac{E_N}{2} - \frac{E_N^3}{2\,m_N^2} +
      \frac{E_N^2}{4\,m_N} - \frac{m_N}{4} \right) \, ,\\[3ex]
   \Pi_V^{00}(\Gamma^n, \vec q) &=& 0 \, ,\\[3ex]
   \Pi_V^{kk}(\Gamma^0, \vec q) &=&  A_{20}\,C\,\left(
  \frac{E_N}{8} + \frac{m_N}{8} + \frac{q_k^2}{4\,m_N} \right)  +
B_{20}\,C\,\left( -\frac{E_N^2}{16\,m_N} + \frac{m_N}{16} - \frac{q_k^2\,E_N}{8\,m_N^2} +
      \frac{q_k^2}{8\,m_N} \right)  \nonumber \\[2.15ex]
&+&  C_{20}\,C\,\left( -\frac{E_N^2}{4\,m_N} + \frac{m_N}{4} + \frac{q_k^2\,E_N}{2\,m_N^2} +
      \frac{q_k^2}{2\,m_N} \right)  \, ,\\[3ex]
   \Pi_V^{kk}(\Gamma^n, \vec q) &=&  A_{20}\,C\,\left(i\,\frac{\epsilon_{k\,n\,0\,\rho}\,
        q_k\,q_\rho }{4\,m_N}\right) +
   B_{20}\,C\,\left( i\,\frac{\epsilon_{k\,n\,0\,\rho}\,
        q_k\,q_\rho }{4\,m_N}\right)\, ,\\[3ex]
   \Pi_V^{k0}(\Gamma^0, \vec q) &=&  A_{20}\,C\,\left(-i\,\frac{q_k}{4} -i\,\frac{q_k\,E_N}{4\,m_N} \right)+
  B_{20}\,C\,\left(-i\,\frac{q_k}{8} +i\, \frac{q_k\,E_N^2}{8\,m_N^2}  \right)+
  C_{20}\,C\,\left(i\,\frac{q_k}{2}-i\, \frac{q_k\,E_N^2}{2\,m_N^2}  \right)\, ,\\[3ex]
   \Pi_V^{k 0}(\Gamma^n, \vec q) &=&  A_{20}\,C\,\left(\,\epsilon_{k\,0\,n\,\rho}\,\left(\frac{
         q_\rho}{8} +\frac{q_\rho\,E_N}{8\,m_N}  \right) \right) +
  B_{20}\,C\,\left(\,\epsilon_{k\,0\,n\,\rho}\,\left(\frac{q_\rho}{8} +
      \frac{q_\rho\,E_N}
       {8\,m_N}   \right)\right) \, ,\\[3ex]
   \Pi_V^{kj}(\Gamma^0, \vec q) &=&    A_{20}\,C\,\frac{q_k\,q_j}
    {4\,m_N} + B_{20}\,C\,\left(-\frac{
           q_k\,q_j \,E_N}{8\,m_N^2} +
      \frac{q_k\,q_j}{8\,m_N} \right)  +
    C_{20}\,C\,\left( \frac{q_k\,
         q_j\,E_N}{2\,m_N^2} +
      \frac{q_k\,q_j}{2\,m_N} \right) \, ,\\[3ex]
   \Pi_V^{kj}(\Gamma^n, \vec q) &=&  A_{20}\,C\,\left(i\,\frac{\epsilon_{k\,n\,0\,\rho}\,
        q_j\,q_\rho }{8\,m_N}+i\,\frac{\epsilon_{j\,n\,0\,\rho}\,
        q_k\,q_\rho }{8\,m_N}\right) +
   B_{20}\,C\,\left(i\,\frac{\epsilon_{k\,n\,0\,\rho}\,
        q_j\,q_\rho }{8\,m_N}+i\,\frac{\epsilon_{j\,n\,0\,\rho}\,
        q_k\,q_\rho }{8\,m_N}\right)\, .\\[3ex]
{\rm {\bf{\underline{Axial-Vector:}}}}\\[3ex]
   \Pi_A^{\mu\nu}(\Gamma^0, \vec q) &=&   0 \, ,\\[3ex]
   \Pi_A^{k0}(\Gamma^n, \vec q) &=&  \tilde{A}_{20}\,C\,\left(\,-i\, \delta_{n\,k}\,
       \left( \frac{E_N}{4} + \frac{E_N^2}{8\,m_N} + \frac{m_N}{8} \right)  -
      i\,\frac{q_k\,q_n}{8\,m_N} \right) +
   \tilde{B}_{20}\,C\,\left(i\,\frac{q_k\,q_n \,E_N}{8\,m_N^2}  \right) \, ,\\[3ex]
   \Pi_A^{kj}(\Gamma^n, \vec q) &=& \tilde{A}_{20}\,C\,\left(\delta_{n\,j}\,
       \left(\frac{q_k }{8} +
         \frac{q_k\,E_N}{8\,m_N} \right)  
      + \delta_{n\,k}\,\left(\frac{q_j }
          {8} +\frac{q_j\,E_N}{8\,m_N} \right)  \right) +
 \tilde{B}_{20}\,C\,\left(-\frac{q_k\,q_j\,q_n}{8\,m_N^2} \right) \, ,
\end{eqnarray}
where $C=\sqrt{2m_N^2 / (E_N(E_N+m_N))}$, $E_N^2=m_N^2 +\vec q^2 $ and the Latin indices $k,j,n$ denote
spatial directions $1,2,3$ and $k\neq j$. A summation is implied over the index $\rho$.

\newpage

{\bf{Appendix B: Chiral perturbation theory results}}

\vspace*{0.5cm}

For convenience we collect in this appendix the results of HB$\chi$PT results taken from Ref.~\cite{Diehl:2006js} for the isovector ($I=1$) and isoscalar ($I=0$)
first moments and axial charge:

\beq
	\tilde{A}_{20}^{I=1}(0)&=&\tilde{A}_{20}^{I=1(0)}\left\lbrace 1 - \frac{m_\pi^2}{(4\pi f_\pi)^2}\left[(2g_A^2+1)\ln\frac{m_\pi^2}{\lambda^2} + g_A^2 \right]\right\rbrace + \tilde{A}_{20}^{I=1(2,m)} m_\pi^2 
\nonumber \\
\tilde{B}_{20}^{I=1}(0)&=&\tilde{B}_{20}^{I=1(0)}\left\lbrace 1 - \frac{m_\pi^2}{(4\pi f_\pi)^2}\left[(2g_A^2+1)\ln\frac{m_\pi^2}{\lambda^2} + g_A^2 \right]\right\rbrace +\tilde{A}_{20}^{I=1(0)} \frac{m_\pi^2g_A^2}{3(4\pi f_\pi)^2}\ln\frac{m_\pi^2}{\lambda^2} + \tilde{B}_{20}^{I=1(2,m)} m_\pi^2 
\eeq

\be
\tilde{A}_{10}^{I=1}(0)=\tilde{A}_{10}^{I=1(0)}\left\lbrace 1 - \frac{m_\pi^2}{(4\pi f_\pi)^2}\left[(2g_A^2+1)\ln\frac{m_\pi^2}{\lambda^2} + g_A^2 \right]\right\rbrace + \tilde{A}_{10}^{I=1(2,m)} m_\pi^2 
\ee

\beq
	A_{20}^{I=1}(0)&=&A_{20}^{I=1(0)}\left\lbrace 1 - \frac{m_\pi^2}{(4\pi f_\pi)^2}\left[(3g_A^2+1)\ln\frac{m_\pi^2}{\lambda^2} + 2g_A^2 \right]\right\rbrace + A_{20}^{I=1(2,m)} m_\pi^2 
\nonumber \\
	B_{20}^{I=1}(0)&=&B_{20}^{I=1(0)}\left\lbrace 1 - \frac{m_\pi^2}{(4\pi f_\pi)^2}\left[(2g_A^2+1)\ln\frac{m_\pi^2}{\lambda^2} + 2g_A^2 \right]\right\rbrace +A_{20}^{I=1(0)} \frac{m_\pi^2g_A^2}{(4\pi f_\pi)^2}\ln\frac{m_\pi^2}{\lambda^2} + B_{20}^{I=1(2,m)} m_\pi^2 
\eeq

\beq
	A_{20}^{I=0}(0)&=&A_{20}^{I=0(0)} + A_{20}^{I=0(2,m)} m_\pi^2\nonumber \\
	B_{20}^{I=0}(0)&=&B_{20}^{I=0(0)}\left[ 1 - \frac{3g_A^2m_\pi^2}{(4\pi f_\pi)^2}\ln\frac{m_\pi^2}{\lambda^2} \right] - A_{20}^{I=0(0)}\frac{3g_A^2m_\pi^2}{(4\pi f_\pi)^2}\ln\frac{m_\pi^2}{\lambda^2} + B_{20}^{I=0(2,m)} m_\pi^2 + B_{20}^{I=0(2,\pi)}(0)  
\eeq

\be
	\tilde{A}_{10}^{I=0}(0)=\tilde{A}_{10}^{I=0(0)}\left\lbrace 1 - \frac{3g_A^2m_\pi^2}{(4\pi f_\pi)^2}\left[\ln\frac{m_\pi^2}{\lambda^2} + 1 \right]\right\rbrace + \tilde{A}_{10}^{I=0(2,m)} m_\pi^2 
\ee

We note that the expressions for $A_{20}^{I=1}$ and $\tilde{A}_{20}^{I=1}$ are
the same as those given in Eq.~(\ref{HBchPT}) (up to a redefinition of $C$, $\tilde{C}$ and $c_8$, $\tilde{c}_8$). We have included them here using
the notation of Ref.~\cite{Diehl:2006js} for completeness.

We performed a combined fit to the following  CB$\chi$PT results taken from Ref.~\cite{Dorati:2007bk}:

\beq
	A_{20}^{I=1}(0) &= &a_{20}^v + \frac{a_{20}^vm_\pi^2}{(4\pi f_\pi)^2} \left[ -(3g_A^2+1)\ln\frac{m_\pi^2}{\lambda^2} - 2g_A^2+g_A^2\frac{m_\pi^2}{M_0^2} \left(1+3\ln\frac{m_\pi^2}{M_0^2}\right) \right.  
\nonumber \\
	&-&\left. \frac{1}{2}g_A^2\frac{m_\pi^4}{M_0^4}\ln\frac{m_\pi^2}{M_0^2}+g_A^2\frac{m_\pi}{\sqrt{4M_0^2-m_\pi^2}}\left(14-8\frac{m_\pi^2}{M_0^2}+\frac{m_\pi^4}{M_0^4}\right) arccos\left(\frac{m_\pi}{2M_0} \right)   \right] 
\nonumber \\
&+& \frac{\Delta a_{20}^v(0)g_Am_\pi^2}{3(4\pi f_\pi)^2} \left[ 2\frac{m_\pi^2}{M_0^2} \left(1+3\ln\frac{m_\pi^2}{M_0^2} \right) - \frac{m_\pi^4}{M_0^4}\ln\frac{m_\pi^2}{M_0^2} + \frac{2m_\pi(4M_0^2-m_\pi^2)^{\frac{3}{2}}}{M_0^4}arccos\left(\frac{m_\pi}{2M_0}\right) \right]
\nonumber \\
&+& 4m_\pi^2\frac{c_8^{(\lambda)}}{M_0^2} + \mathcal{O}(p^3) \\
B_{20}^{I=1}(0) &=& b_{20}^v\frac{M_N(m_\pi)}{M_0} + \frac{a_{20}^vg_A^2m_\pi^2}{(4\pi f_\pi)^2}\left[\left(3+\ln\frac{m_\pi^2}{M_0^2}\right)-\frac{m_\pi^2}{M_0^2}\left(2+3\ln\frac{m_\pi^2}{M_0^2}\right)   \right.	\nonumber \\
	& +& \left.\frac{m_\pi^4}{M_0^4}\ln\frac{m_\pi^2}{M_0^2} -\frac{2m_\pi}{\sqrt{4M_0^2-m_\pi^2}}\left(5-5\frac{m_\pi^2}{M_0^2}+\frac{m_\pi^4}{M_0^4}\right)       arccos\left(\frac{m_\pi}{2M_0}\right) \right]+ \mathcal{O}(p^3) \\
C_{20}^{I=1}(0) &=& c_{20}^v\frac{M_N(m_\pi)}{M_0} + \frac{a_{20}^vg_A^2m_\pi^2}{12(4\pi f_\pi)^2}\left[ -1 + 2\frac{m_\pi^2}{M_0^2}\left(1+\ln\frac{m_\pi^2}{M_0^2}\right) \right.	\nonumber \\
	& -& \left.\frac{m_\pi^4}{M_0^4}\ln\frac{m_\pi^2}{M_0^2} +\frac{2m_\pi}{\sqrt{4M_0^2-m_\pi^2}}\left(2-4\frac{m_\pi^2}{M_0^2}+\frac{m_\pi^4}{M_0^4}\right)       arccos\left(\frac{m_\pi}{2M_0}\right) \right]+ \mathcal{O}(p^3)
\eeq

\beq
	A_{20}^{I=0}(0) &=& a_{20}^s + 4m_\pi^2\frac{c_9}{M_0^2} - \frac{3a_{20}^s g_A^2 m_\pi^2}{(4\pi f_\pi)^2} \left[\frac{m_\pi^2}{M_0^2} +  \frac{m_\pi^2}{M_0^2}\left(2-\frac{m_\pi^2}{M_0^2}\right)\ln\frac{m_\pi}{M_0} \right.  
\nonumber \\
	&+&\left.\frac{m_\pi}{\sqrt{4M_0^2-m_\pi^2}}\left(2-4\frac{m_\pi^2}{M_0^2}+\frac{m_\pi^4}{M_0^4}\right) arccos\left(\frac{m_\pi}{2M_0} \right)   \right] + \mathcal{O}(p^3)
\\
B_{20}^{I=0}(0) &=& b_{20}^s\frac{M_N(m_\pi)}{M_0} - \frac{3a_{20}^sg_A^2m_\pi^2}{(4\pi f_\pi)^2}\left[\left(3+\ln\frac{m_\pi^2}{M_0^2}\right)-\frac{m_\pi^2}{M_0^2}\left(2+3\ln\frac{m_\pi^2}{M_0^2}\right)   \right.	
\nonumber \\
&+&	\left. \frac{m_\pi^4}{M_0^4}\ln\frac{m_\pi^2}{M_0^2} -\frac{2m_\pi}{\sqrt{4M_0^2-m_\pi^2}}\left(5-5\frac{m_\pi^2}{M_0^2}+\frac{m_\pi^4}{M_0^4}\right)       arccos\left(\frac{m_\pi}{2M_0}\right) \right]+ \mathcal{O}(p^3)
\\
C_{20}^{I=0}(0) &= &c_{20}^s\frac{M_N(m_\pi)}{M_0} - \frac{a_{20}^sg_A^2m_\pi^2}{4(4\pi f_\pi)^2}\left[ -1 + 2\frac{m_\pi^2}{M_0^2}\left(1+\ln\frac{m_\pi^2}{M_0^2}\right) \right.	\nonumber \\
	&-&\left. \frac{m_\pi^4}{M_0^4}\ln\frac{m_\pi^2}{M_0^2} +\frac{2m_\pi}{\sqrt{4M_0^2-m_\pi^2}}\left(2-4\frac{m_\pi^2}{M_0^2}+\frac{m_\pi^4}{M_0^4}\right)       arccos\left(\frac{m_\pi}{2M_0}\right) \right]+ \mathcal{O}(p^3)
\eeq

where $M_0$ is the mass of the nucleon at the chiral limit.

\newpage

{\bf{Appendix C: Numerical results for the isovector sector}}

\vspace*{0.5cm}

\begin{table}[h]
\begin{center}
\begin{tabular}{|c|c|c|c|c|c|}
\hline\hline
 $m_\pi$ (GeV)  & $(Q)^2$  & $A_{20}$ & $B_{20}$ & $\tilde A_{20}$ & $\tilde B_{20}$    \\
 (no. confs)   &   &  &   &  &    \\\hline
\multicolumn{4}{c}{$\beta=3.9$, $24^3\times 48$ }\\\hline
            &0.0	  & 0.256(9)  & 0.364(23) & 0.307(6)   & 0.651(99)   \\
            &0.322	  & 0.230(7)  & 0.337(18) & 0.273(5)   & 0.487(121) \\
   0.4675   &0.619(1)& 0.202(9)  & 0.299(16) & 0.252(7)   & 0.568(72)  \\
  (477)     &0.897(2)& 0.178(12) & 0.277(22) & 0.227(11)  & 0.456(69)  \\
            &1.157(3)& 0.172(19) & 0.249(34) & 0.176(17)  & 0.078(96)  \\
            &1.404(4)& 0.154(22) & 0.232(34) & 0.191(22)  & 0.319(65)  \\
            &1.640(6)& 0.136(40) & 0.208(60) & 0.193(51)  & 0.346(126) \\
 \hline
            &0.0	   & 0.257(10)  & 0.418(34)& 0.310(7)   & 0.516(109) \\
            &0.321	   & 0.219(8)   & 0.361(29)& 0.264(7)   & 0.431(155)  \\
0.4319      &0.615(1) & 0.185(9)   & 0.296(26)& 0.232(9)   & 0.378(75)  \\
(365)       &0.888(3) & 0.167(13)  & 0.257(34)& 0.211(14)  & 0.339(86)   \\
            &1.143(4) & 0.168(26)  & 0.209(42)& 0.175(25)  & 0.191(115) \\
            &1.385(6) & 0.141(29)  & 0.151(42)& 0.173(37)  & 0.242(99)  \\
            &1.614(8) & 0.103(48)  & 0.118(63)& 0.136(67)  & 0.139(112) \\
 \hline
        &0.0	  & 0.258(10) & 0.408(44)  & 0.296(8)  &  0.683(150)   \\
        &0.320	  & 0.217(10) & 0.360(32)  & 0.266(7)  &  0.878(170)    \\
0.3770  &0.613(1)& 0.192(11) & 0.338(26)  & 0.253(8)  &  0.417(89)     \\
  (553) &0.884(3)& 0.164(20) & 0.265(40)  & 0.237(26) &  0.349(132)    \\
        &1.138(4)& 0.204(76) & 0.339(129) & 0.288(91) &  0.661(301)     \\
        &1.377(6)& 0.166(59) & 0.271(97)  & 0.218(70) &  0.402(176)   \\
        &1.604(8)& 0.105(92) & 0.140(126) & 0.143(113)&  0.215(210)    \\
\hline
       &0.0	   & 0.255(16)  & 0.309(40) & 0.284(10)  & 0.277(136)     \\
       &0.317(1)   & 0.239(13)  & 0.241(36) & 0.268(10)  & 0.380(184)     \\
 0.3032&0.601(2)   & 0.199(13)  & 0.268(32) & 0.223(11)  & 0.048(104)     \\
 (943) &0.862(4)   & 0.174(23)  & 0.178(37) & 0.195(20)  & 0.245(110)     \\
       &1.103(6)   & 0.135(27)  & 0.156(45) & 0.172(24)  & 0.304(118)     \\
       &1.330(8)   & 0.098(21)  & 0.142(34) & 0.135(22)  & 0.180(70)      \\
       &1.543(10)  & 0.105(34)  & 0.097(40) & 0.087(27)  & 0.051(75)      \\
\hline
\end{tabular}
\caption{Results on $A_{20}$ and $B_{20}$ form factors at $\beta=3.9$,
lattice size: $24^3\times48$}
\label{tab:results 3.9a}
\end{center}
\end{table}

\begin{table}[h]
\begin{center}
\begin{tabular}{|c|c|c|c|c|c|}
\hline\hline
 $m_\pi$ (GeV)  & $(Q)^2$  & $A_{20}$ & $B_{20}$ & $\tilde A_{20}$ & $\tilde B_{20}$    \\
 (no. confs)   &   &  &   &  &    \\\hline
\multicolumn{4}{c}{$\beta=3.9$, $32^3\times 64$ }\\\hline
      & 0.0	    & 0.243(15)  & 0.375(67) & 0.287(10)  & 0.413(95)   \\
      & 0.183	    & 0.230(14)  & 0.382(67) & 0.262(8)   & 0.454(215)  \\
0.2978& 0.354(1)  & 0.207(14)  & 0.318(54) & 0.244(7)   & 0.282(98)    \\
 (351)& 0.516(1)  & 0.196(14)  & 0.298(50) & 0.228(9)   & 0.298(100)   \\
      & 0.670(2)  & 0.164(20)  & 0.249(61) & 0.228(15)  & 0.441(138)   \\
      & 0.817(3)  & 0.165(16)  & 0.206(44) & 0.204(11)  & 0.318(65)    \\
      & 0.957(4)  & 0.154(18)  & 0.188(40) & 0.179(14)  & 0.165(64)    \\
      & 1.222(6)  & 0.145(28)  & 0.173(55) & 0.191(38)  & 0.252(110)   \\
      & 1.348(7)  & 0.095(23)  & 0.170(42) & 0.151(36)  & 0.123(81)    \\
\hline
       &0.0	    & 0.263(13)  & 0.301(47) & 0.275(13)  & 0.752(174)     \\
       &0.182	    & 0.240(09)  & 0.284(52) & 0.261(10)  & 0.666(331)     \\
0.2600 &0.352(1)  & 0.222(10)  & 0.252(38) & 0.259(11)  & 0.582(164)     \\
 (667) &0.512(1)  & 0.196(12)  & 0.221(40) & 0.256(16)  & 0.524(158)     \\
       &0.664(2)  & 0.185(17)  & 0.246(48) & 0.216(19)  & 0.263(185)     \\
       &0.808(3)  & 0.161(14)  & 0.233(34) & 0.217(16)  & 0.378(105)    \\
       &0.945(4)  & 0.143(15)  & 0.204(34) & 0.181(18)  & 0.247(93)      \\
       &1.205(6)  & 0.100(22)  & 0.122(42) & 0.127(24)  & 0.140(128)     \\
       &1.328(7)  & 0.106(22)  & 0.162(40) & 0.124(26)  & 0.072(105)    \\
\hline
\end{tabular}
\caption{Results on $A_{20}$ and $B_{20}$ form factors at $\beta=3.9$,
lattice size: $32^3\times64$}
\label{tab:results 3.9b}
\end{center}
\end{table}

\begin{table}[h]
\begin{center}
\begin{tabular}{|c|c|c|c|c|c|}
\hline\hline
 $m_\pi$ (GeV)  & $(Q)^2$  & $A_{20}$ & $B_{20}$ & $\tilde A_{20}$ & $\tilde B_{20}$   \\
 (no. confs)   &   &  &   &  &   \\\hline
\multicolumn{4}{c}{$\beta=4.05$, $32^3\times 64$ }\\\hline
      &0.0	    & 0.258(9)     & 0.431(26) &  0.303(7)     & 0.579(92)  \\
      &0.294	    & 0.231(7)     & 0.382(22) &  0.270(6)     & 0.458(166)\\
0.4653&0.568(1)  & 0.210(7)     & 0.329(20) &  0.248(7)     & 0.523(73) \\
 (419)&0.824(2)  & 0.197(12)    & 0.283(25) &  0.232(11)    & 0.342(77) \\
      &1.067(3)  & 0.170(19)    & 0.313(41) &  0.221(20)    & 0.263(116)\\
      &1.297(4)  & 0.166(19)    & 0.256(30) &  0.211(21)    & 0.367(75) \\
      &1.517(5)  & 0.156(31)    & 0.208(43) & 0.181(33)    & 0.256(76)  \\
      &1.930(7)  & 0.084(29)    & 0.085(34) & 0.111(30)    & 0.190(92)  \\
      &2.126(9)  & 0.056(58)    & 0.054(63) & 0.066(65)    & 0.114(139) \\
\hline
      &0.0	      & 0.244(12)  &  0.465(46) & 0.312(10)  & 0.625(114)   \\
      &0.293	      & 0.240(10)  &  0.434(42) & 0.287(10)  & 0.455(206)   \\
0.4032&0.564(1)    & 0.208(11)  &  0.356(36) & 0.263(11)  & 0.477(91)    \\
 (326)&0.816(2)    & 0.197(17)  &  0.302(43) & 0.259(19)  & 0.539(132)   \\
      &1.053(3)    & 0.175(32)  &  0.225(48) & 0.251(34)  & 0.307(154)   \\
      &1.278(5)    & 0.144(20)  &  0.203(36) & 0.195(22)  & 0.332(81)    \\
      &1.493(6)    & 0.144(41)  &  0.204(66) & 0.202(56)  & 0.387(137)   \\
      &1.895(9)    & 0.068(25)  &  0.095(36) & 0.093(26)  & 0.142(73)   \\
      &2.084(10)   & 0.043(26)  &  0.083(45) & 0.071(32)  & 0.151(76)    \\
\hline
      &0.0(0)      & 0.231(24)  &  0.426(67) & 0.310(23)  & 0.283(268) \\
      &0.291(1)    & 0.237(20)  &  0.307(66) & 0.279(15)  & 0.328(322) \\
0.2925&0.556(2)    & 0.216(21)  &  0.343(47) & 0.252(18)  & 0.459(202) \\
 (447)&0.801(3)    & 0.236(41)  &  0.279(69) & 0.223(33)  & 0.025(207) \\
      &1.029(5)    & 0.136(39)  &  0.116(77) & 0.165(34)  & 0.028(238) \\
      &1.245(7)    & 0.139(72)  &  0.090(84) & 0.219(95)  & 0.484(266) \\
      &1.450(10)   & 0.104(42)  &  0.132(67) & 0.157(52)  & 0.273(129) \\
\hline
\end{tabular}
\caption{Results on $A_{20}$ and $B_{20}$ form factors at $\beta=4.05$}
\label{tab:results 4.05}
\end{center}
\end{table}

\begin{table}[h]
\begin{center}
\begin{tabular}{|c|c|c|c|c|c|}
\hline\hline
  $m_\pi$ (GeV)  & $(Q)^2$  & $A_{20}$ & $B_{20}$ & $\tilde A_{20}$ & $\tilde B_{20}$   \\
 (no. confs)   &   &  &   &  &   \\\hline
\multicolumn{4}{c}{$\beta=4.2$, $32^3\times 64$ }\\\hline
      &0.0	  & 0.252(13)  & 0.402(35) & 0.295(9)   & 0.768(126)  \\
      &0.467(1)   & 0.215(12)  & 0.343(29) & 0.272(8)   & 0.606(145)  \\
0.4698&0.886(2)   & 0.205(14)  & 0.264(29) & 0.260(11)  & 0.595(95)   \\
(357) &1.270(4)   & 0.214(27)  & 0.220(47) & 0.249(27)  & 0.523(141)  \\
      &1.627(6)   & 0.140(37)  & 0.240(78) & 0.178(47)  & 0.030(159)  \\
      &1.961(8)   & 0.131(36)  & 0.127(40) & 0.177(39)  & 0.226(113)  \\
      &2.276(10)  & 0.118(29)  & 0.057(25) & 0.114(28)  & 0.146(83)   \\
      &2.861(14)  & 0.056(40)  & 0.017(43) & 0.054(28)  & 0.058(100)  \\
\hline
\multicolumn{4}{c}{$\beta=4.2$, $48^3\times 96$}\\ \hline
       & 0.0	     &0.276(21)  & 0.234(45) &0.290(15)  &  0.579(129)  \\
       & 0.211	     &0.255(14)  & 0.264(49) &0.275(12)  &  0.301(375) \\
 0.2622& 0.407(2)    &0.253(16)  & 0.205(38) &0.279(11)  &  0.689(146) \\
  (245)& 0.589(3)    &0.228(17)  & 0.157(45) &0.255(12)  &  0.268(137) \\
       & 0.762(5)    &0.184(30)  & 0.180(59) &0.272(27)  &  0.453(213) \\
       & 0.925(7)    &0.193(23)  & 0.178(38) &0.231(15)  &  0.177(107) \\
       & 1.081(9)    &0.174(22)  & 0.146(39) &0.230(17)  &  0.207(100) \\
       & 1.373(13)   &0.133(39)  & 0.076(49) &0.167(23)  &  0.168(114)\\
       & 1.511(16)   &0.093(28)  & 0.142(40) &0.164(20)  &  0.189(82)  \\
       & 1.644(18)   &0.054(38)  & 0.153(52) &0.118(29)  & -0.001(95)  \\
       & 1.773(20)   &0.026(37)  & 0.138(62) &0.152(40)  &  0.031(107) \\
\hline
\end{tabular}
\caption{Results on $A_{20}$ and $B_{20}$ form factors at $\beta=4.2$ }
\label{tab:results 4.2}
\end{center}
\end{table}

\newpage

\end{document}